\documentclass[aps,prb,twocolumn,showpacs]{revtex4}

\usepackage{amsmath}
\usepackage{epsfig}
\newcommand{\bea}{\begin{eqnarray}}
\newcommand{\eea}{\end{eqnarray}}
\newcommand{\be}{\begin{equation}}
\newcommand{\ee}{\end{equation}}

\begin{document}

\title{Coulomb drag between two spin incoherent Luttinger liquids}

\author{Gregory A. Fiete,$^1$ Karyn Le Hur,$^2$ and Leon Balents$^3$}

\affiliation{$^1$Kavli Institute for Theoretical Physics, University of California, Santa Barbara, CA 93106, USA\\$^2$D\'epartment de Physique and RQMP, Universit\'e de Sherbrooke, Sherbrooke, Qu\'ebec, Canada J1K 2R1
\\$^3$Physics Department, University of California, Santa Barbara, CA 93106, USA}

\begin{abstract}

In a one dimensional electron gas at low enough density,
the magnetic (spin) exchange energy $J$ between neighboring electrons
is exponentially suppressed relative to the characteristic charge
energy, the Fermi energy $E_F$.  At non-zero temperature $T$, the energy
hierarchy $J \ll T \ll E_F$ can be reached, and we refer to this as
the spin incoherent Lutinger liquid state. We discuss the Coulomb drag
between two parallel quantum wires in the spin incoherent regime, as
well as the crossover to this state from the low temperature regime by
using a model of a fluctuating Wigner solid.  As the temperature
increases from zero to above $J$ for a fixed electron density, the
$2k_F$ oscillations in the density-density correlations are lost.  As
a result, the temperature dependence of the Coulomb drag is
dramatically altered and non-monotonic dependence may result.  Drag between 
wires of equal and unequal density are discussed, as well as the effects of weak 
disorder in the wires.  We speculate that weak disorder may play an important
role in extracting information about quantum wires in real drag experiments.

\end{abstract}

\date{\today}
\maketitle



\section{Introduction}
\label{sec:intro}

In recent years correlated electron systems at the nanoscale and in
reduced dimensions have attracted much
attention.\cite{Manoharan:nat00,Nygard:nat00,Jarillo:nat05} In one
spatial dimension electron correlations are expected to be enhanced,
leading to the so called Luttinger liquid (LL)
state.\cite{haldane81,Voit:rpp95} The existence of the LL liquid state
in a one dimensional electron system is now an established
experimental fact,\cite{ishii:nat03,Bockrath:nat99,Yao:nat99} with
direct measurements of the distinct spin and charge velocities in
momentum resolved tunneling (as predicted by the theory) providing
compelling evidence.\cite{yacoby:sci02,Steinberg:sci05}

\begin{figure}[t]
\includegraphics[width=.6\linewidth,clip=]{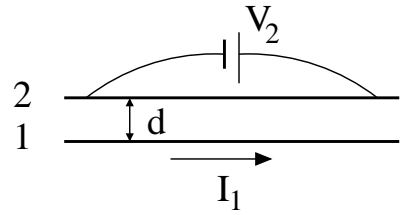}
\caption{\label{fig:drag} Coulomb drag schematic.  Two quantum wires are arranged parallel to one another.  A dc current $I_1$ flows in the ``active'' wire 1 and a voltage bias $V_2$ is measured in the ``passive'' wire 2 when $I_2=0$.}
\end{figure}

Another way to explore the correlations in one dimensional systems is
in a drag experiment between two parallel quantum wires or
nanotubes.\cite{Yamamoto:pe02,Debray:jpcm01,Debray:sst02} (Recall that
in most cases the bare conductance of a quantum wire is $2e^2/h$ per
transverse channel regardless of the electron
interactions,\cite{Maslov:prb95,Safi:prb99} and so does not reveal
information about electron correlations. An exception to this
quantization condition is the situation discussed by Matveev in
Refs.~\onlinecite{Matveev:prl04,Matveev:prb04}.)  The typical drag
set-up involves a current driven in a ``active'' wire while a voltage
drop is measured in a ``passive'' wire. See Fig.~\ref{fig:drag}.

The quantity often taken to
describe the drag effect is the ``drag resistivity'' (drag per unit length)
defined as 
\be 
\label{eq:drag_formula}
r_D = -\lim_{I_1\to 0} \frac{e^2}{h}\frac{1}{L} \frac{dV_2}{dI_1}\;, 
\ee 
where $V_2$ is a voltage induced in wire 2 (the ``passive'' wire) due
to a current $I_1$ in wire 1 (the ``active'' wire).  Here $e$ is the
charge of the electron, $h$ Planck's constant, and $L$ the length of
the wire.  The sign of the drag can be either positive or negative,
but it is generally positive (note minus sign in formula) for
repulsive interactions between electrons.

Typically, one is interested in the dependence of $r_D$ on the
temperature, the interwire spacing, the electron density and Fermi
wave vector in each wire, the disorder, the wire length, and possibly
on an external magnetic field.  Physically drag is the result of
collisions (momentum transfer) from electrons in the active wire which
tend to ``push'' or ``pull'' the electrons in the passive wire.
Electrons in the passive wire move under these collisions until an
electric field is built up in the passive wire (due to a non-uniform
density of electrons there) which just cancels the force of the
momentum transfer of the electrons in the active wire.  This is the
physics behind the well known drag formulas of Zheng and
MacDonald,\cite{Zheng:prb93} and Pustilnik {\em et
al.}\cite{Pustilnik:prl03}  (See Eq.~\eqref{eq:drag_Pustilnik}.)

While the experimental data on drag between quantum wires is
limited,\cite{Yamamoto:pe02,Debray:jpcm01,Debray:sst02} a fair amount
of theoretical work has been published.  Various studies have made use
of LL
theory,\cite{Fuchs:prb05,Klesse:prb00,Nazarov:prl98,Schlottmann:prb04,Flensberg:prl98}
Fermi liquid (FL) theory with\cite{Raichev:prb00} and
without\cite{Pustilnik:prl03} multiple sub-bands, effects of
inter-wire tunneling,\cite{Raichev:prl99} effects of
disorder,\cite{Ponomarenko:prl00} shot noise
correlations,\cite{Gurevich:prb00,Trauzettel:prl02} mesoscopic
fluctuations,\cite{Mortensen:prb02,Mortensen:prl01} and the effects of
different signs of electron exchange interactions in the
wires.\cite{Schlottmann_2:prb04} Additionally, phonon mediated drag
has been studied in one dimension.\cite{Muradov:prb02,Raichev:prb01}
The main qualitative difference that is found between the LL and FL
approaches is whether $r_D$ tends to increase (LL) or decrease (FL) as
the temperature is reduced to the lowest values. For the case of two
infinitely long indentical clean wires the following results are
obtained: In a LL, electrons tend to ``lock'' into a commensurate
state at the lowest tempertures giving rise to a diverging drag, while
in a FL the phase space available for scattering tends to zero as the
temperature is lowered, implying a vanishing drag.  For two
non-identical wires the drag is usually significantly suppressed at
low temperatures relative to the drag in the identical
case.\cite{Pustilnik:prl03,Fuchs:prb05}

In spite of the theoretical effort, a number of open questions remain.
In particular, the drag effect is known to be strongest when the
electron density is low,
\cite{Yamamoto:pe02,Debray:jpcm01,Debray:sst02} which typically
implies that electron interactions are strong, or equivalently that
$r_s \equiv a/(2a_B)$ is large with $a=n^{-1}$ and $a_B$ the Bohr radius.  For
very strong interactions, there is an exponential separation of the
spin exchange energy, $J$, and the characteristic charge energy, $E_F$, which at
finite temperatures can lead to incoherent (thermally excited) spin degrees of freedom
while the charge degrees of freedom remain approximately coherent and
close to their ground state.\cite{Matveev:prl04,Matveev:prb04}
This energy hierarchy at finite temperature $T$, $J \ll T \ll E_F$, we
refer to as the spin incoherent Luttinger liquid regime.  Already there
is mounting understanding of how such spin incoherent Luttinger
liquids behave in the Green's
function,\cite{cheianov03,Cheianov04,Fiete:prl04} in the momentum
distribution function,\cite{Cheianov:cm04} in momentum resolved
tunneling,\cite{Fiete:prb05} and in
transport.\cite{Matveev:prl04,Matveev:prb04,Fiete_2:prb05,Kindermann:prl05}

Our goal in this work is to explore some of the implications of spin
incoherence on the drag between two quantum wires.  We consider only
the simplest case of a single channel wire.  We attempt to elucidate
what qualitative and quantitative changes one can expect for the
Coulomb drag when the temperature is much smaller or larger than $J$.
Since $J/E_F$ is expontially small,\cite{Matveev:prl04,Matveev:prb04} 
a small change in the temperature
can induce a dramatic change in the temperature dependence
of the drag.  Based on earlier work of the
authors,\cite{Fiete_2:prb05} we are able to discuss the drag deep in the
spin incoherent regime in terms of spinless electrons using a simple
mapping between the charge variables of the charge sector of a LL with
spin and the variables of a spinless LL.  

\begin{figure}[t]
\includegraphics[width=.9\linewidth,clip=]{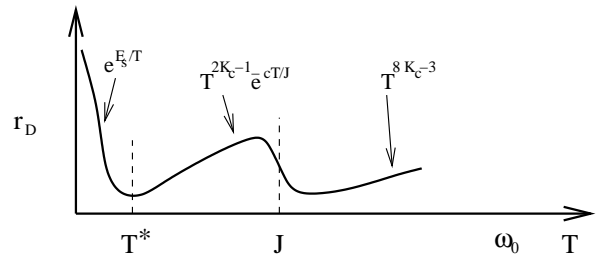}
\caption{\label{fig:temp_regimes} Schematic of the possible temperature dependence of the Coulomb drag in a wire of sufficiently low electron density that $J \ll E_F \lesssim \omega_0$.  The non-monotonic temperature dependence shown can be obtained for $K_c>1/2$ and $\tilde U(4k_F) \ll \tilde U(2k_F)$ which may be realized for widely space wires.  The
temperature $T^*$ is the ``locking" temperature of two identical wires, below which the drag exhibits activated behavior and $E_s$ is an energy gap associated with the ``locking".  In this paper we consider only temperatures $T>T^*$ and investigate the limits $T <J$ and $T>J$, as well as the crossover between the two.  When $J\ll E_F$ a sharp drop in the drag resistance should be observable for $T \sim J$.}
\end{figure}

The crossover to the spin
incoherent regime is discussed using a model of a fluctuating Wigner
solid with an antiferromagnetic Heisenberg spin chain in the spin
sector. Distortions of the solid couple the spin and charge degrees of
freedom. The model allows us to quantitatively address the crossover.
The main result is that as the temperature increases from zero to
above $J$ for a fixed electron density, the (already weak) $2k_F$
oscillations in the density-density correlations are rapidly lost.  As a
result, the drag is dramatically suppressed (when $k_Fd \gtrsim 1$) 
since forward scattering contributions vanish in the LL model and the Wigner solid model, and
$4k_F$ contributions are suppressed relative to $2k_F$ contributions
by $\tilde U(4k_F)/\tilde U(2k_F)\sim e^{-2k_Fd} \;{\rm for}\;
k_Fd \gtrsim 1$.  Here $\tilde U(k d)$ is the Fourier transform of the interwire
electron-electron interaction, $d$ is the interwire separation, and
$k_F\equiv \pi/(2a)$ is the Fermi wavevector.  See Fig.~\ref{fig:temp_regimes}.

In addition to a possible dramatic suppression (depending on various physical parameters) of the
drag, a non-monotonic dependence on temperature may also occur, as illustrated schematically in
Fig.~\ref{fig:temp_regimes}.  Specifically, when $T>T^*$, where $T^*$ is the ``locking" temperature (see Sec.~\ref{sec:general} below),
we find that the temperature dependence of the drag is given by
\be
r_D=C_1T^{\alpha_{2k_F}}f(T/J)+C_2 T^{\alpha_{4k_F}},
\ee
where $f(X\to0)\to1$ and $f(X\gtrsim1)\sim e^{-c X}$, with $c$ a
constant of order one.  The coefficients $C_1 \propto \tilde U(2kF)$ and
$C_2 \propto \tilde U(4kF)$.  The exponents $\alpha_{2k_F/4k_F}$ depend on the interactions in the wires and on the presence or absence of disorder.
We find
\bea 
\alpha_{2k_F}&=&2K_c-1\hspace{1 cm}{\rm clean},\nonumber \\
&=&  2K_c \hspace{1 cm} {\rm disordered}\;,
\eea
and
\bea 
\alpha_{4k_F}&=&8K_c-3\hspace{1.5 cm}{\rm clean},\nonumber \\
&=&  8K_c-2 \hspace{1 cm} {\rm disordered}\;.
\eea

While it is interesting that disorder changes the temperature dependence
of the drag, it may play an even more important role in the measurement
of the drag effect itself in actual experiments.  The reason is that the
drag effect is maximal for identical wires; any $k_F$ mismatch results
in a drag that is generally strongly diminished relative to this case,
indeed, {\sl exponentially} so at low temperature.  Disorder tends to
``smear'' the momentum structure of the density response that determines
the drag, eliminating this exponential suppression.  Hence, since in a
real experiment one will never have truly identical wires, some residual
disorder may actually play a key role in experimental studies of drag
between one dimensional systems.

This paper is organized in the following way. In
Sec.~\ref{sec:general} we discuss general considerations for drag
between two quantum wires with an emphasis on features relevant to the
effects of being in the spin incoherent regime.  In
Sec.~\ref{sec:Wigner} we discuss a model of a fluctuating Wigner
solid with a Heisenberg spin chain to describe the magnetic (spin) degrees of
freedom.  We derive expressions for the density fluctuations of the electron gas by
including magneto-elastic coupling that induces $2k_F$ density modulations
at low temperatures ($T \ll J$) and results in an instability towards a local
lattice distortion favoring a spin-Peierls-like state. In Sec.~\ref{sec:spectral} we 
discuss the temperature dependence of the  Fourier transform of the 
density-density correlation function in detail.  In Sec.~\ref{sec:drag} we discuss the
drag itself in detail by considering different temperature regimes, the effects of disorder,
and the case of wires with mismatched density.  In Sec.~\ref{sec:summary} 
we summarize the main results of our paper and
in the appendicies we give some exact formulas for the density-density
correlation function relevant to the spin coherent-incoherent
crossover as well as other useful formulas.

\section{General Considerations}
\label{sec:general}

We assume the Hamiltonian of our system is of the form
\be
\label{eq:H}
H=H_1+H_2+H_{12}\;,
\ee
 where $H_i$ is the Hamiltonian in the $i^{th}$ wire and $H_{12}$ describes
the interactions between electrons in different wires.  A proper drag
situation is one in which the tunneling between wires can be
neglected.  We thus assume that $H_{12}$ allows only interactions
which forbid electrons to tunnel between the wires.\cite{tunneling_comment}  The Hamiltonian
$H_i$ in principle describes arbitrary interactions between electrons
within the $i^{th}$ wire; depending on the particular situation of
interest, a number of different models have been proposed from Fermi
liquid\cite{Debray:sst02,Raichev:prb00,Pustilnik:prl03} to Luttinger
liquid.\cite{Fuchs:prb05,Klesse:prb00,Nazarov:prl98,Schlottmann:prb04,Flensberg:prl98}

\begin{figure}[h]
\includegraphics[width=.8\linewidth,clip=]{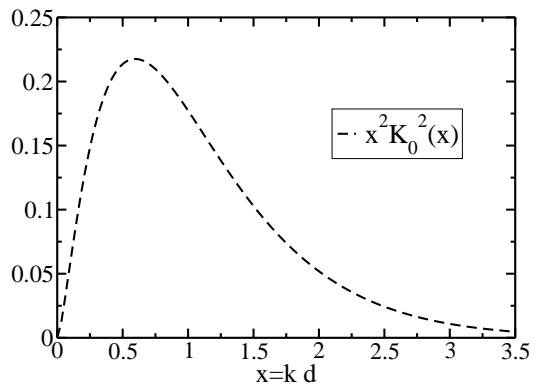}
\caption{\label{fig:potential} Momentum dependence of the the quantity $k^2  \tilde U^2_{12}(k)=\left(\frac{2 e^2}{\epsilon d}\right)^2 (x K_0(x))^2$ appearing in the drag formula Eq.~\eqref{eq:drag_Pustilnik} as a function of the dimensionless variable $x= k d$ where we have assumed the real space inter-wire electron interaction \eqref{eq:U}.  Plotted is the quantity 
$(x K_0(x))^2$ which has a maximum for $x \approx 0.6.$  For $x \ll 1$, $(x K_0(x))^2\sim (-\gamma +\ln(2)-\ln(x))^2x^2$ where $\gamma \approx 0.5772$ is Euler's constant, while for $x\gg 1$, $(x K_0(x))^2\sim e^{-2x}.$}
\end{figure}

The Hamiltonian $H_{12}$ is a function of the interwire electron
interaction which is often taken to be of the form
\be
\label{eq:U}
U_{12}(x)=\frac{e^2}{\epsilon}\frac{1}{\sqrt{d^2+x^2}}\;,
\ee
where $e$ is the charge of the electron, $\epsilon$ the dielectric
constant of the material, and $d$ the separation between the two
wires. The Fourier transform, $\tilde U(k)=2(e^2/\epsilon)K_0(k d)$, 
depends on the 
dimensionless parameter $kd$ and for $kd \gg 1,\; \tilde U(k)\sim \frac{1}{\sqrt{k}}e^{-kd}$.
This exponential dependence of the interwire interaction leads to an exponential dependence
of the drag resistivity on wire separation (for large enough $d$) when the
drag is dominated by large momentum transfer as it is in the LL model and the
model we will discuss in this paper. The following drag formula 
(or its equivalent) has been derived by Zheng and MacDonald\cite{Zheng:prb93}
for a disordered FL,\cite{2Ddrag} by Klesse and Stern\cite{Klesse:prb00} for a LL, and
by Pustilnik {\em et  al.}\cite{Pustilnik:prl03} for a clean FL:
\be 
\label{eq:drag_Pustilnik}
r_D=\int_0^\infty dk \int_0^\infty d\omega \frac{k^2  \tilde U^2_{12}(k)}{4 \pi^2 n_1 n_2 T} \frac{{\rm Im}\chi^R_1(k,\omega) {\rm Im}\chi^R_2(k,\omega)}{\sinh^2(\omega/2T)}\;,
\ee
where ${\rm Im}\chi^R_i(k,\omega)$ is the imaginary part of the Fourier transformed 
retarded density-density correlation function, Eqs.~\eqref{eq:chi_t} and \eqref{eq:chi_tau},
and $n_i$ is the density of the 
electrons in the $i^{th}$ wire.  Knowing the general features of the
interwire interaction $U_{12}$, (see Fig.~\ref{fig:potential}) it is clear that to determine the drag
one must determine the ${\rm Im}\chi^R_i(k,\omega)$. 
 We will discuss the behavior of
 ${\rm Im}\chi^R_i(k,\omega)$ and its temperature dependence 
in detail in Sec.~\ref{sec:spectral}, but for now
it is sufficient to point out that most of weight occurs near momenta
of $k \approx 0,\;k\approx 2k_F$, and $k \approx 4k_F$ so that one may write
\bea
\label{eq:chi}
{\rm Im}\chi^R_i(k,\omega) \approx {\rm Im}\chi_i^{R,0}(k,\omega)+{\rm Im}\chi_i^{R,2k_F}(k,\omega)\nonumber \\
+{\rm Im}\chi_i^{R,4k_F}(k,\omega).
\eea
 As we will
see, the $2k_F$ components of ${\rm Im}\chi^R_i(k,\omega)$ present at low temperatures,
$T \ll J$, will disappear when $T \gg J$.  Moreover, once it is known that
${\rm Im}\chi_i^{R,0}(k,\omega) \propto \delta(\omega-v_{c,i}k)$ (see Sec.~\ref{sec:spectral})
where $v_{c,i}$ is the charge velocity of the $i^{th}$ wire, it is readily seen
from Eq.~\eqref{eq:drag_Pustilnik} that for the LL model (or any model
possessing a harmonic bosonized theory) the $k=0$ contribution to the drag resistivity
vanishes (in the absence of disorder), leaving only contributions from 
the higher $k=2k_F,4k_F$ values of  ${\rm Im}\chi^R_i(k,\omega)$.  The higher 
$k=2k_F,4k_F$ values can be strongly suppressed by the interwire interaction because 
$\tilde U^2_{12}(k)\sim e^{-2kd}$ when $kd \gg 1$.  Hence, in the limit $k_F d \gtrsim 1$
we expect a dramatic reduction in the drag and a change in the temperature dependence
of the drag (see below) when $2k_F$ oscillations are lost and only $4k_F$
modulations remain.  However, before we address these details, it is useful to go over what we can
say about drag deep in the spin incoherent regime itself.

Based on the results of earlier work,\cite{Fuchs:prb05,Klesse:prb00}
we can immediatly discuss the drag deep within the spin incoherent LL
regime, $J \ll T \ll E_F$.  It has earlier been
argued\cite{Fiete_2:prb05} that the spin incoherent LL behaves
essentially as a spinless LL by noting that the Hamiltonian is
diagonal in spin to lowest order in $J/T \ll 1$, and by demanding the
equivalence of the physical charge density in both cases. The
equivalence can be summarized by the simple equation $K=2K_c$ (in the
notation of Klesse and Stern\cite{Klesse:prb00}) relating the
interaction parameter of the spinless LL theory and the interaction
parameter of the charge sector of the LL theory with spin. (Strictly speaking,
for drag the relevant interaction parameter is
$K=K_-$, the interaction parameter in the odd channel of the relative density
of the two wires,\cite{Fuchs:prb05,Klesse:prb00} but when the inter-wire interactions
are sufficiently weak $K_-$ is essentially equal to the correponding parameter 
of the isolated wire.)  The relation $K=2K_c$ is valid for any particle conserving
operator;\cite{Fiete_2:prb05} the drag resistivity is derived from
such an operator. Hence, those general results apply here.

As discussed in Refs.~\onlinecite{Fuchs:prb05,Klesse:prb00}, drag in
the spinless LL model comes from the backscattering of electrons.  As
long as the inter-wire interactions are sufficiently small the
effects of backscattering can be treated perturbatively.  Based on
such a perturbative treatment, Klesse and Stern
found\cite{Klesse:prb00} that for identical wires there is a
temperature scale $T^*$ that separates high and low temperature drag
regimes.  (It is assumed througout this paper that $T^*$ is greater than the
thermal length $T_L$ so that the Fermi liquid leads attached to the
quantum wire are not felt.)  For temperatures $T \gg T^*,T_L$, the
drag varies as a power of the temperature, while for temperatures $
T^* \gg T \gg T_L$, the drag resistivity shows activated behavior with
a gap of order $T^*$ itself. The physics is similar to that of a pinned
charge density wave. This result follows from an analysis of a sine-Gordon model
in the odd channel of the coupled wire problem.  Applying the equivalence rule
 $K=2K_c$ discussed above in the spin incoherent regime, we have
\be
\label{eq:rhoD_high}
\rho_D \sim T^{8K_{c}-3}, \;\; T \gg T^*,T_L,J,
\ee
\be
\label{eq:rhoD_low}
\rho_D \sim e^{E_s/T}, \;\; T^* \gg T \gg T_L,J,
\ee
where $E_s\sim
T^*$.\cite{Klesse:prb00} Note that for $3/8 <
K_{c} < 3/4$ the temperature dependence of the spinless (fully
polarized) electron gas and the spin incoherent electron gas exhibit
very different drag resistivity behavior with temperature when
$T>>T^*$. (The spinless case has a diverging drag resistivity as $T$
is lowered, while the spin incoherent case has a suppressed drag
resistivity as $T$ is lowered.)  This is qualitatively similar to the
transport results found in Ref.~\onlinecite{Fiete_2:prb05} for $1/2
<g_c <1$.

The results \eqref{eq:rhoD_high} and \eqref{eq:rhoD_low} above were
derived from a perturbative analysis of the sine-Gordon equation
which results from treating the backscattering in LL theory.  For most
realistic parameter values, the baskcattering strength flows to strong
coupling and the resulting state is that of the two quantum wires
locked into a ``zig-zag'' charge pattern.  The value of $T^*$
depends on details of the quantum wire system such as the
density, wire widths, and separation $d$,\cite{Klesse:prb00} but 
for most realistic situations $T^*/E_F \ll 0.01$.

It is interesting to consider how spin incoherence affects the
``zig-zag'' locking pattern of the electrons in the two wires. 
The relative size of $J$ and $T^*$ will
determine what the periodicity of the ``zig-zag'' pattern will be for
$T <<T^*$.  For $J \ll T \ll T^*$, there is a ``$4k_F$" locking (seen
easily from the $K=2K_c$ mapping\cite{Fiete_2:prb05}) since
$T \gg J$ ensures $2k_F$ pieces of the density are washed out, while
for $T \ll J, T^*$, there is a ``$2k_F$" locking.  Of course, for $T^*
\ll J\ll T$ the locking phase is not obtained.  Throughout this paper we will
assume $T \gg T^*$ so that we need not be concerned with
``locking" from here forward.

Similar arguments to those given in Ref.~\onlinecite{Fuchs:prb05} can
also be used to describe the incommensurate-commensurate transition
deep in the spin incoherent regime for wires of different electron densities.
We now leave generalities behind and turn to a detailed calculation of the
drag itself in the regime of very strongly interacting one dimensional electrons. 

\section{The fluctuating Wigner Solid Model}
\label{sec:Wigner}

We assume from the outset that the interactions between the electrons are very strong, which 
typically means the density is low enough that we can treat the electrons in
each wire as a harmonic chain\cite{Novikov} in the charge sector and a nearest
neighbor Heisenberg antiferromagnet in the spin sector:\cite{Matveev:prb04,Fiete:prb05}

\begin{equation}
H_{\rm wire}=H_c+H_s\;,
\end{equation}
where 
\begin{equation}
\label{eq:H_c}
H_c=\sum_{l=1}^N  \frac{p_l^2}{2m}+\frac{m\omega_0^2}{2}(u_{l+1}-u_l)^2\;,
\end{equation}
is the Hamiltonian in the charge sector with $p_l$ the momentum of the
$l^{th}$ electron, $u_l$ the displacement from equilibrium of the
$l^{th}$ electron, $m$ the electron mass, and $\omega_0$ the frequency of 
local electron displacements (this will depend on the electron density, 
the width of the wires, the dielectric constant of the material, and other
parameters such as the distance to a nearby
gate\cite{Glazman:prb92,Hausler:prb02}).  The position of the electrons 
along the chain are given by
\begin{equation}
x_l=l a +u_l\;,
\end{equation}
where $a$ is the mean spacing of the electrons.  The Hamiltonian of the spin sector takes the form
\begin{equation}
\label{eq:H_s}
H_s=\sum_l J_l {\vec S_{l+1}}\cdot{\vec S_l}\;.
\end{equation}
Note that in (\ref{eq:H_s}) the coupling $J_l$ between spins depends on the distance between them.  Assuming that the fluctuations from the equilibrium positions are small compared to the mean
particle spacing, we can expand the exchange energy as
\begin{equation}
J_l=J_0 +J_1(u_{l+1}-u_l) + {\cal O}((u_{l+1}-u_l) ^2)\;.
\end{equation}
In this case the full Hamiltonian takes the form
\begin{equation}
\label{eq:H_wire}
H=H_c + H_s + H_{s-c}\;,
\end{equation}
where
\begin{equation}
\label{eq:H_m-e}
 H_{s-c}= J_1\sum_l (u_{l+1}-u_l){\vec S_{l+1}}\cdot{\vec S_l} \;.
\end{equation}
Here $H_{s-c}$ represents a magneto-elastic coupling as it couples the
magnetic modes to the elastic distortions of the lattice that constitute
the charge modes.

Our goal is to evaluate the Fourier transform of the retarded density-density correlation function
$-i\theta(t-t')\langle [\rho(x,t),\rho(x',t')]\rangle$ [which
appears in the drag formula \eqref{eq:drag_Pustilnik}] up to second order in $J_1$ for
$T,J_0 \ll \hbar \omega_0$, for both $T\ll J_0$ and $T \gg J_0$.  
We use the following definition of the electron density:
$\rho(x,t)=\sum_l \delta(x-al -u_l(t))$.  An exact calculation within
this model is presented in Appendix~\ref{app:dens_exact}.  Here we will pursue an
approximate calculation that captures all of the essential features of
the more exact perturbative results.

\subsection{Low energy approach to charge fluctuations}

In this work we are concerned only with energies (temperatures) small
compared to the characteristic lattice energy, i.e. $T \ll \hbar
\omega_0$, but still large compared to the ``locking" temperature $T^*$.  
When $T \ll J_0$ further
approximations that can be made, but for now our only
restriction will be that $T \ll \hbar \omega_0$.  We begin by
expanding the displacement of the electron density in a Fourier
series.  For low energy distortions the $k\approx 0$ component is most
important, while the magneto-elastic term \eqref{eq:H_m-e}
couples the $k \approx \pi/a$ component to the spin operator ${\vec
S_{l+1}}\cdot{\vec S_l}$. Thus,
the displacement, $u_l$ of the $l^{th}$ electron in the harmonic chain 
\eqref{eq:H_c} is approximately given by
\be
\label{eq:u}
u_l=u_0(la)+u_\pi(la)(-1)^l,
\ee
where $u_0$ refers the  $k\approx 0$ component of the displacement and 
$u_\pi$ refers to the $k\approx \pi/a$ displacement.  Both $u_0$ and $u_\pi$
are assumed to be slowly varying functions of position, and we expect
$u_\pi \ll u_0$. 

\subsubsection{Low energy form of the action}

\begin{figure}[h]
\includegraphics[width=.9\linewidth,clip=]{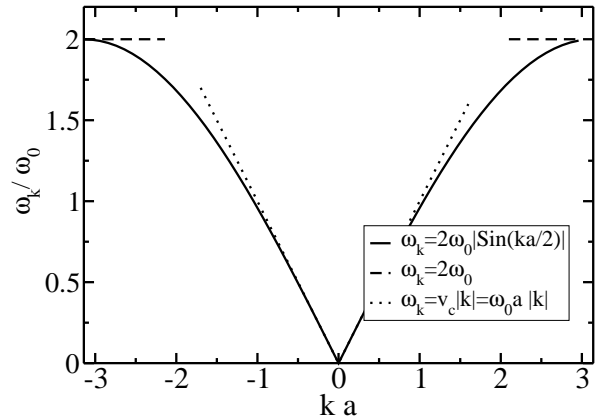}
\caption{\label{fig:spectrum} Approximate form for the phonon spectrum.
There are two relevant parts of the spectrum that enter the action $S_c$, the part near $k\approx 0$ and the part near $k\approx \pi/a$.  The phonons with $k\approx 0$ are described by a Debye model while those phonons with $k \approx \pi/a$ are described by an Einstein model.
}
\end{figure}

The action for the low density electron gas is
\be
\label{eq:S}
S=S_c+S_s+S_{s-c}\,.
\ee
Using the expression \eqref{eq:u} for the particle displacement
fields, and noting that from the phonon dispersion of \eqref{eq:H_c}, 
$\omega_k=2\omega_0|\sin(ka/2)|$, one has for $k\approx 0$ the dispersion $\omega_k=v_c
|k|$ with $v_c = \omega_0 a$, while for $k\approx \pi/a$ the dispersion
$\omega_k = 2 \omega_0$ is independent of $k$. (See Fig.~\ref{fig:spectrum}.)
In the charge sector we then
have the following action of the form $S_c=S_c^0+S_c^\pi$,
\bea
\label{eq:S_c}
S_c =\int \frac{dx d\tau}{2\pi}\Biggl[\frac{1}{K_c v_c}\left((\partial_\tau \theta_c(x,\tau))^2+v_c^2 (\partial_x \theta_c(x,\tau))^2\right)\nonumber \\
+ m\pi \left((\partial_\tau u_\pi(x,\tau))^2 + (2\omega_0)^2 u_\pi(x,\tau)^2\right) \Biggr]\;,\hspace{.5 cm}
\eea
where $K_c=\pi \hbar/(2amv_c)$. Note the lack of spatial derivative in the 
$u_\pi$ piece of the action which results in the absence of any $k$ dependence 
in $\omega_k$ near $k\approx \pi/a$. In effect, we have described the small $k$
oscillations (phonons) with a Debye-type model and the large $k$ oscillations as an
Einstein model.  Fig.~\ref{fig:spectrum} illustrates the approximations to the
full phonon spectrum.

In the standard LL model for weakly interacting electrons $K_c=v_F/v_c\lesssim1$ where
$v_F$ is the Fermi velocity of non-interacting electrons. In the present case of strongly
interacting electrons we have $K_c=\frac{\hbar^2\pi}{2ma^2}\frac{1}{\hbar \omega_0}\sim \frac{E_F}{\hbar \omega_0}\sim \frac{\rm{Kin}}{{\rm Pot}}\sim \frac{1}{r_s}$ where $E_F$ is the Fermi energy of non-interacting electrons and $r_s\equiv a/(2a_B)$ where $a_B=\epsilon \hbar^2/me^2$ is the Bohr radius
of a material of dielectric constant $\epsilon$. 
Thus, in the strongly interacting limit $K_c$ scales roughly as the ratio of the kinetic energy to
the potential energy which itself roughly scales as $r_s^{-1}$ implying that very strong (long-range) interactions can lead to small $K_c$.  In practice, however, $K_c$ rarely appears to be smaller than
$0.2$ or  so.

By making the identification ${\cal E}_l\equiv (-1)^l {\vec S_{l+1}}\cdot{\vec S_l}\to {\cal E}(x,\tau)$ in the continuum limit, we have spin-charge coupling in the action
\be
S_{s-c}=\int \frac{dx d\tau}{2\pi} 2 J_1 u_\pi(x,\tau){\cal E}(x,\tau)\;.
\ee

In the special situation where $T \ll J_0$, the action for the spin sector \eqref{eq:H_s} can be bosonized
as\cite{Subir,Giamarchi}
\be
\label{eq:S_s}
S_s=\int \frac{dx d\tau}{2\pi}\Biggl[\frac{1}{K_s v_s}\left((\partial_\tau \theta_s(x,\tau))^2+v_s^2 (\partial_x \theta_s(x,\tau))^2\right)\Biggr]\;,
\ee
where the $SU(2)$ symmetry of the Heisenberg model implies that 
$K_s =1$ and the spin velocity is $v_s\sim  J_0 a/\hbar$.  
(Here intrawire backscattering effects in the spin sector have been neglected 
and a sine-Gordon term dropped. Since we are ultimately interested in energies/temperatures
much larger than $J_0$ or much smaller than $J_0$ this will not affect any of our conclusions.)
However, when $T \gtrsim J_0$ the action for the spin sector must describe more accurately the
short distance physics of the Heisenberg chain.  Nevertheless, the action \eqref{eq:S_s} will 
prove useful in understanding the approach to the spin incoherent regime in the limit
$T \to J_0$ from below.

Finally, for $T \ll J_0$ the coupling term in the action can be expressed as\cite{Giamarchi}
\be
\label{eq:S_sc}
S_{s-c}=\int \frac{dx d\tau}{2\pi} \frac{2 J_1}{2\pi \alpha a} u_\pi(x,\tau) \sin(\sqrt{2}\theta_s(x,\tau))\;,
\ee
where we have used the low-energy bosonized form $(-1)^l {\vec S_{l+1}}\cdot{\vec S_l} \approx \frac{1}{2\pi \alpha}\sin(\sqrt{2}\theta_s(x))$.  Here $\alpha={\cal O}(a)$ is a short distance cut off of order the lattice spacing.  At low energies this term will lead to the spin-Peierls state indicated in Fig.~\ref{fig:Wigner_solid}.

\subsubsection{Expressing the density}

Expanding the density and making use of \eqref{eq:u} then  gives (we have suppressed
the time dependence of $u_0$ and $u_\pi$ immediately below for clarity of presentation)
\bea
\label{eq:rho_u}
\rho(x) &=& \sum_l \delta(x-la-u_l),\nonumber \\
&=&\sum_l \delta(x-la-u_0(la)+u_\pi(la)(-1)^l),\nonumber \\
&\approx& \sum_l [\delta(x-la-u_0(la))\nonumber \\
&& \;\;\;\;-\delta'(x-la-u_0(la))u_\pi(la)(-1)^l)].
\eea
Multiplying by $\delta(x'-la)$, integrating over $x'$, and 
using the Poisson summation identity,
\be
\sum_l \delta(x'-la)=\frac{1}{a}\sum_{m=-\infty}^{\infty}e^{i \frac{2\pi}{a}m x'},
\ee
the expression for the density becomes
\bea
\rho(x)\approx \int dx'\frac{1}{a}\delta(x-x'+u_0(x')) \sum_{m=-\infty}^{\infty}e^{i \frac{2\pi}{a}m x'}\nonumber \\
\times \Bigl[1+\cos(\frac{\pi}{a}x')\partial_{x'}u_\pi(x')\nonumber \\
+ i\frac{\pi}{2a}\left((2m+1)e^{i\frac{\pi}{a}x'}+(2m-1)e^{-i\frac{\pi}{a}x'}\right)u_\pi(x')\Bigr],
\eea
where we have integrated by parts in the second term of \eqref{eq:rho_u}.  
Performing
the integration over $x'$, making use of the approximate relation 
$\delta(x-x'-u_0(x'))\approx \delta(x'-x-u_0(x))/|1+\partial_xu_0(x)|$, 
and assuming $a\partial_x u_\pi(x) \ll u_\pi(x)$, we find the most 
important terms up to $4k_F$ are
\bea
\label{eq:rho_eff}
\rho(x)\approx \frac{1}{a}(1-\partial_xu_0(x))\Bigl[1-\frac{2\pi}{a}\sin\left(2k_F(x+u_0(x))\right)u_\pi(x)\nonumber \\
+\cos\left(4k_F(x+u_0(x))\right)\Bigr],\nonumber \\
\approx \rho_0 -\frac{\sqrt{2}}{\pi}\partial_x\theta_c(x)-\rho_0 \frac{2\pi}{a}\sin\left(2k_Fx+\sqrt{2}\theta_c(x)\right)u_\pi(x)\nonumber \\
+ \rho_0 \cos\left(4k_Fx+\sqrt{8}\theta_c(x)\right), \hspace{.5 cm}
\eea
where $\rho_0 \equiv 1/a$, $k_F \equiv \pi/(2a)$, and we have made the 
identification $u_0(x)/a=\sqrt{2}\theta_c(x)/\pi$. Recall the field $\theta_c$ is governed by
the action \eqref{eq:S_c}.  This formula 
resembles the standard bosonized expression for the density of a 
Luttinger liquid (except for the term $u_\pi$ multiplying the $2k_F$ 
part of the density instead of a term involving the spin fields).  As we will
now see, a formula very similar to that obtained from the standard 
Luttinger liquid treatment results from integrating out the
high energy $u_\pi$ phonon modes in favor of the low energy spin modes.  
To see this, consider only the $2k_F$ part of the density and compute $\langle \rho_{2k_F}(x,\tau)\rangle$
to lowest order in the action $S_{s-c}$ and integrate out the $u_\pi$ fields to obtain a new $\rho_{2k_F}^{\rm eff}(x,\tau)$
independent of $u_\pi$.  At lowest order we find
\bea
\label{eq:upi_first}
\langle u_\pi(x,\tau)\rangle = -\frac{2J_1}{\hbar} \int \frac{dx'd\tau'}{2\pi} {\cal E}(x',\tau')\nonumber \\
\times \langle T_\tau u_\pi(x,\tau) u_\pi(x',\tau')\rangle_{S_c^\pi}\;,
\eea
where $T_\tau$ is the $\tau$-ordering operator and the $\tau$-ordered product is evaluated in
the action $S_c^\pi$ given in \eqref{eq:S_c}. The $\tau$-ordered product is readily evaluated as
\bea
\label{eq:uu}
\langle u_\pi(x,\tau) u_\pi(x',\tau')\rangle_{S_c^\pi}=
\frac{\hbar \delta(x-x')}{4m\omega_0}\nonumber \\
\times \left[\frac{e^{2\omega_0|\tau-\tau'|}}{e^{\beta2\omega_0}-1}-\frac{e^{-2\omega_0|\tau-\tau'|}}{e^{-\beta2\omega_0}-1}\right]\;,\nonumber \\
\to \frac{\hbar \delta(x-x')}{4m\omega_0} e^{-2\omega_0|\tau-\tau'|}
\;{\rm as}\; \beta \omega_0\to \infty.
\eea
Recall that we are interested only in temperatures low compared to the phonon energy
$\omega_0$ so the limit $\beta \omega_0\to \infty$ is the appropriate one. Here $\beta=(k_B T)^{-1}$
where $k_B$ is Boltzmann's constant.  The integral over
position in \eqref{eq:upi_first} is immediately evaluated with the delta function 
in \eqref{eq:uu} and the remaining
integral over $\tau'$ can be approximately evaluated under the assumption $\beta \omega_0\to \infty$
which produces the dominant contribution at $\tau' \approx \tau$ with a width in $\tau'$ of order $1/(2\omega_0)$ resulting in
\be
\langle u_\pi(x,\tau)\rangle \approx -\frac{2J_1}{16\pi^2 m \omega_0^2} {\cal E}(x,\tau)\;,
\ee
which yields
\bea
\label{eq:r_2kF_eff}
\rho_{2k_F}^{\rm eff}(x,\tau) = \frac{1}{8\pi}\left(\frac{J_1}{m\omega_0^2 a^2}\right)
\sin\left(2k_Fx+\sqrt{2}\theta_c(x,\tau)\right)\nonumber \\
\times {\cal E}(x,\tau).\;\;\;\
\eea

The result \eqref{eq:r_2kF_eff} is general,  and  valid whenever $T \ll \hbar \omega_0$.
However, when $T \ll J_0$ one may use the expression ${\cal E}(x,\tau)=\frac{1}{2\pi \alpha}\sin(\sqrt{2}\theta_s(x,\tau))$ which leads to the familiar looking density 
 \bea
 \label{eq:rho_eff_final}
\rho^{\rm eff}(x,\tau) = \rho_0 -\frac{\sqrt{2}}{\pi}\partial_x\theta_c(x,\tau)\hspace{3.5 cm}\nonumber \\ - 
\frac{\rho_0}{16\pi}\left(\frac{J_1}{m\omega_0^2 a^2}\right)
\sin\left(2k_Fx+\sqrt{2}\theta_c(x,\tau)\right)\sin(\sqrt{2}\theta_s(x,\tau))
\nonumber \\
+ \rho_0 \cos\left(4k_Fx+\sqrt{8}\theta_c(x,\tau)\right).\hspace{.5 cm}
\eea

The expression for the effective density \eqref{eq:rho_eff_final} with the high energy $u_\pi$ 
modes integrated out in favor of the spin variables is valid for $T \ll J_0$ and may
be compared directly with the LL result obtained for weakly interacting electrons.\cite{Voit:rpp95}   
At temperatures $T \gg J_0$ \eqref{eq:r_2kF_eff} must be used for the $2k_F$ density
variations.
The only material difference between \eqref{eq:rho_eff_final} and the standard LL result is the dimensionless ratio of spin and charge energies $\left(\frac{J_1}{m\omega_0^2 a^2}\right)\sim\frac{v_s}{v_c}$ which is absent (since it is of order 1) in the familiar 
LL case.  When the interactions are strong as we have assumed them to be here, then 
$\left(\frac{J_1}{m\omega_0^2 a^2}\right)\sim\frac{v_s}{v_c} \ll 1$, since $J_1$ diminishes and $\omega_0$ increases 
with increasing strength of the interactions.  However, starting from the strongly interacting limit
and decreasing the interaction strength
the ratio $\left(\frac{J_1}{m\omega_0^2 a^2}\right) \to 1$.  It is worth emphasizing, then, that
when the temperature is low compared to both spin and charge energies a 1-d
electron gas always behaves as a LL in the sense of the various power laws that will appear
in the correlation functions, although the overall prefactors of the $2k_F$ pieces will be
down by the ratio $\left(\frac{J_1}{m\omega_0^2 a^2}\right)$.  If, on the other hand, 
the system is in the spin incoherent regime $J_0 \ll T \ll \hbar \omega_0$, the $2k_F$ parts of the 
correlations will be washed out from thermal effects.  We now turn to an investigation
of how this happens in detail for the case of the density-density correlation function
and then discuss the implications for the Coulomb drag between two quantum wires.

\section{Evaluation of ${\rm Im}\chi^R_i(k,\omega)$}
\label{sec:spectral}

Here we consider two limits of the double wire system shown in Fig.~\ref{fig:drag}: (i)
Clean wires without disorder and (ii) Wires with weak disorder.  The case of strong disorder
is uninteresting as the electrons are all localized over the relevant energy/length scales
of the experiment.  As we discussed in Sec.~\ref{sec:general}, the drag formula
\eqref{eq:drag_Pustilnik} generically contains contributions at $k \approx 0,\;k\approx 2k_F$, and $k \approx 4k_F$ so that 
${\rm Im}\chi^R_i(k,\omega) \approx {\rm Im}\chi_i^{R,0}(k,\omega)+{\rm Im}\chi_i^{R,2k_F}(k,\omega)+{\rm Im}\chi_i^{R,4k_F}(k,\omega)$.  
We now turn to an evaluation of each of these pieces.  We have used the two standard 
(equivalent) definitions 
\bea
\label{eq:chi_t}
\chi^R_i(k,\omega)=-i \int_{-\infty}^\infty \!\!\! dx  \int_{0}^\infty \!\!\! dt e^{i((\omega +i\eta)t -k x)}
\nonumber \\
\times \langle [(\rho_i(x,t)-\rho_{i,0}), (\rho_i(0,0)-\rho_{i,0})]\rangle,
\eea
and
\bea
\label{eq:chi_tau}
\chi_i(k,\omega_n)=-\int_{-\infty}^\infty \!\!\! dx  \int_{0}^\beta \!\!\! d\tau e^{i(\omega_n \tau -k x)}
\nonumber \\
\times \langle (\rho_i(x,\tau)-\rho_{i,0}) (\rho_i(0,0)-\rho_{i,0})\rangle,
\eea
where $\rho_{i,0}$ is the average density of the $i^{\rm th}$ wire, and $\eta$ is a small 
infinitesimal that ensures convergence of the time integral in \eqref{eq:chi_t}.  The retarded
correlation function is obtained from \eqref{eq:chi_tau} via the substitution $i\omega_n\to
\omega+i\eta$.  We will use both of the formulas above in the subsections that follow.

\subsection{Clean Wires}

We first consider wires with no disorder.  We will also assume initially that $T \ll J_0$ so that 
we may use the form of the density \eqref{eq:rho_eff_final}. (This is only an issue for
the evaluation of ${\rm Im}\chi_i^{R,2k_F}(k,\omega)$ since ${\rm Im}\chi_i^{R,0}(k,\omega)$ and ${\rm Im}\chi_i^{R,4k_F}(k,\omega)$ do not involve the spin sector of the Hamiltonian.) As the authors discussed in Ref.~\onlinecite{Fiete_2:prb05} the approach to the spin incoherent regime from temperatures well below the spin energy can be understood in this way.  In all calculations below,
recall that we have assumed the temperature is low, $T \ll \hbar \omega_0$, so that the charge
sector is always in the LL regime and described by the action $S_c^0$ in \eqref{eq:S_c}.

\subsubsection{${\rm Im}\chi_i^{R,0}(k,\omega)$}

We first evaluate the $k\approx 0$ piece of the retarded density-density correlation function. 
From the expression \eqref{eq:rho_eff_final}, we have 
\be
\label{eq:rho_0_expr}
\rho^{\rm eff}_0(x,t) = \rho_0 -\frac{\sqrt{2}}{\pi}\partial_x\theta_c(x,t),
\ee
whose correlation function is readily computed (see Appendix~\ref{app:FT}) to yield
\be
\label{eq:chi_R0_final}
{\rm Im}\chi^{R,0}_i(k,\omega)=\frac{k^2}{a^2}\frac{\hbar L}{2m\omega_k}\left[\pi \delta(\omega-\omega_k)-\pi \delta(\omega+\omega_k)\right].
\ee
The equation above, \eqref{eq:chi_R0_final}, is the central result of this subsection and it is worth
pausing to emphasize some of its features.  Most notably, while the calculation was done at finite
temperature, there is no temperature dependence of ${\rm Im}\chi^{R,0}_i(k,\omega)$.  Thus, the
finite temperature $k \approx 0$ response is identical to the zero temperature response.  This means
that temperature does not ``broaden'' the zero temperature $\delta$-function reponse.  Moreover,
for the model at hand, at small $|k|$, $\omega_k=v_{i,c}|k|$ so that for a given $k$ there is a unique value of $\omega_k$.   This means then when the result \eqref{eq:chi_R0_final} is substituted into the
drag formula \eqref{eq:drag_Pustilnik} the drag is identically zero. (An exception is the measure
zero point where the wires are identical, i.e. $v_{1,c}=v_{2,c}$, and the drag response is infinite.
For real wires this precise matching is not possible and the $k=0$ part of the drag generically vanishes.)

Ultimately, the vanishing of the $k \approx 0$ drag is a result of the delta functions appearing in \eqref{eq:chi_R0_final}.  It is expected that the delta functions will be broadened\cite{Pustilnik:prl03,Teber:cm05} in a more complete treatment and that this will lead to a non-zero and temperature
dependent $k \approx 0$ contribution to the drag.

In our work here, we have assumed from the outset that the electron
interactions are very strong and a direct bosonization of the electron
operator is not valid.  Instead, the approximation we have made to
obtain the action \eqref{eq:S_c}, which is formally identical to that
obtained for weakly interacting electrons with a linear dispersion
(aside from the $u_\pi$ terms), is to treat the displacements of
electrons to lowest order in the Taylor series: $(u_{l+1}-u_l)/a \approx
\sqrt{2}\partial_x \theta_c(x)/\pi$.  Including higher derivatives would
result in an interacting bosonic theory and would likewise broaden the
delta functions in \eqref{eq:chi_R0_final} by an amount inversely
proportional to the lifetime and would yield a finite $k \approx 0$
drag.  The precise nature of this contribution to the drag is still a
subject of ongoing research.\cite{Pustilnik:prl03,Teber:cm05}  It is
therefore difficult to compare it quantitatively in theoretical
calculations to the $2k_F$ and $4k_F$ contributions.  However, we expect
that it may be larger or smaller than the latter depending upon
circumstances.  For instance, the $k\approx 0$ drag is clearly {\sl
  subdominant} for drag between identical, clean wires, with repulsive
interactions.  Fortunately, for our purposes of discerning the spin
coherent to incoherent crossover at $T\approx J$, we may satisfy
ourselves with the observation that the $k \approx 0$ drag is in any
case {\sl featureless} at this temperature.  Hence, it can easily be
``subtracted'' by looking for strong temperature-dependent changes in
the drag in this temperature window.

\subsubsection{${\rm Im}\chi_i^{R,2k_F}(k,\omega)$}

The $2k_F$ component of the density response and its temperature dependence is the 
central issue in this paper and we now turn to it in detail.  We have already discussed
general features of the spin incoherent limit $T \gg J_0$ in Sec.~\ref{sec:general},
and we will discuss other more detailed and quantitative features of that regime in the next subsection
where we consider ${\rm Im}\chi_i^{R,4k_F}(k,\omega)$.  Here, we will initially assume
that the temperature is low compared to the spin energies, $T \ll J_0$, and use the
low energy density expression \eqref{eq:rho_eff_final}.  Starting from the low temperature limit 
we show that as the temperature becomes of order the spin energy, the temperature dependence
of the $2k_F$ part of the drag changes and rapidly vanishes as $J_0 \to 0$ for fixed $T \gg  J_0$. 
We also show that in the low temperature limit we recover the temperature dependence of the drag obtained by Klesse and Stern\cite{Klesse:prb00} for electrons with spin.  When $k_F d \gtrsim 1$
the loss of $2k_F$ contributions to the drag (when $T \approx J_0$) implies (via Eq.~\eqref{eq:drag_Pustilnik} and  Fig. \ref{fig:potential}) that there is expected to be a dramatic reduction in the drag over a very small 
temperature window when only the $4k_F$ contribution remains, as  $\tilde U(4k_F)/\tilde U(2k_F)\sim e^{-2k_Fd} \;{\rm for}\; k_Fd \gtrsim 1$.  

The $2k_F$ part of the low energy density operator \eqref{eq:rho_eff_final} is
\bea
\rho^{\rm eff}_{2k_F}(x,t) = -\frac{\rho_0}{16\pi}\left(\frac{J_1}{m\omega_0^2 a^2}\right)
\sin\left(2k_Fx+\sqrt{2}\theta_c(x,\tau)\right)\nonumber \\
\times \sin(\sqrt{2}\theta_s(x,\tau))\;\;\;
\eea
which leads to the following finite temperature result for the $2k_F$ part of the density-density
correlation function computed from \eqref{eq:S_c} and \eqref{eq:S_s}
\bea
\label{eq:rho_2kF_temp}
-i\langle [\rho^{\rm eff}_{2k_F}(x,t),\rho^{\rm
  eff}_{2k_F}(0,0)]\rangle=
\left(\frac{\rho_0}{16\pi}\right)^2\left(\frac{J_1}{m\omega_0^2 a^2}\right)^2 \hspace{2 cm}\nonumber \\
\times \cos(2k_F x) {\rm Im}\Biggl[ \frac{(\pi
  T\alpha/v_c)^{K_c}}{[\sinh\left(\frac{\pi T}{v_c}(x-
    v_ct)\right)\sinh\left(\frac{\pi T}{v_c}(x+
    v_ct)\right)]^\frac{K_c}{2}}\;\;\;\;
\nonumber \\
\times\frac{(\pi T\alpha/v_s)^{K_s}}{[\sinh\left(\frac{\pi T}{v_s}(x-
    v_st)\right)\sinh\left(\frac{\pi T}{v_s}(x+
    v_st)\right)]^\frac{K_s}{2}}\Biggr]. \hspace{1 cm} \eea Here
$\alpha$ is a short-distance cut off of order the lattice spacing.  We
note that in Eq.~(\ref{eq:rho_2kF_temp}) -- and in subsequent similar
formulae -- singularities at $x=\pm v_c t, \pm v_s t$ are regularized by
infinitesimal imaginary parts to the time $t$, which for ease of
presentation are not shown. It is worth pointing out that because of the
hyperbolic nature of the correlation function at finite temperature, a
temperature dependent ``coherence length" naturally appears in both the
spin and charge sectors.  From inspection, the charge coherence length
$\xi_c (T)= v_c/(K_c \pi T)\sim a \hbar \omega_0/(k_B T)$, and the spin
coherence length $\xi_s(T) = v_s/(K_s \pi T)\sim a J_0/(k_B T)$.  Strong
interactions imply $v_s/v_c \ll 1$ ($J_0 \ll \hbar \omega_0$) so that
$\xi_s(T) \ll \xi_c (T)$.  Note that $\xi_s(T) \approx a$ when $T
\approx J_0$.

\begin{figure}[t]
\includegraphics[width=.8\linewidth,clip=]{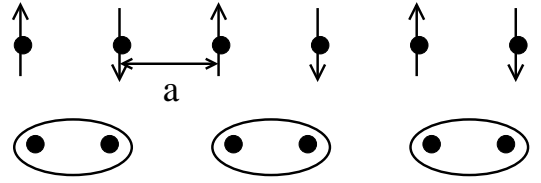}
\caption{\label{fig:Wigner_solid} Classical Wigner solid model.  The
electrons are indicated by solid black dots and are separated by a
mean spacing distance $a$. Top: A ``cartoon'' of the antiferromagnetic
spin arrangement is indicated by the arrows. The lattice has density
oscillations of smallest wavevector $4k_F$.  Bottom: Magneto-elastic
coupling allows the system to lower its energy by slightly distoring
the lattice in order to gain magnetic energy.  Stronger spin
correlations are indicated by the ovals which pair neighboring
electrons at spacing less than $a$.  The lattice has density
oscillations of smallest wavevector $2k_F$, indicating twice the
period of the undistorted lattice.  When $T \gg J_0$ the (small) lattice
distortion is thermally washed out leaving only the $4k_F$ periodicity
of the underlying Wigner solid.}
\end{figure}

Our task is now to substitute \eqref{eq:rho_2kF_temp}  into the integral in \eqref{eq:chi_t} and evaluate the integrals over position and time.  Unfortunately, this integral does not appear to have a closed,
analytical form.  Nevertheless, its general structure is apparent.  At zero temperature the structure in the $(k,\omega)$ plane is very similar to that of the Green's function already computed by Voit\cite{Voit:prb93} and by Meden and Sch\"onhammer.\cite{Meden:prb92}  Depending on the value of $K_c$
there are singularities or thresholds at $\omega=v_sk_\pm$ and  $\omega=v_ck_\pm$,
where $k_\pm=k\pm2k_F$.  With small, but finite temperature these features are smoothed out.  However, as the temperature
increases towards $J_0$ the overall weight in $\chi_i^{R,2k_F}(k_\pm,\omega)$ begins to rapidly
diminish.  To see this, consider the limit $T \to J_0$ from below.  Then $\chi_i^{R,2k_F}(k_\pm,\omega)$
can be bounded as
\bea
\chi_i^{R,2k_F}(k_\pm,\omega) < \left(\frac{k_B T}{J_0}\right)^{K_s} e^{-c\frac{k_B T}{J_0}} 
\!\!\! \int_{-\infty}^\infty \!\!\!\!\!\! d x  \int_{0}^\infty \!\!\!\!\!\! dt e^{i\left(\omega t -k_\pm x\right)}\nonumber \\
\times \left(\frac{\rho_0}{16\pi}\right)^2\left(\frac{J_1}{m\omega_0^2 a^2}\right)^2\hspace{4 cm} \nonumber \\
\times
{\rm Im}\Biggl[\frac{(\pi \alpha T/v_c)^{K_c}}{[\sinh\left(\frac{\pi T}{v_c}(x- v_ct)\right)\sinh\left(\frac{\pi T}{v_c}(x+ v_ct)\right)]^\frac{K_c}{2}}\Biggr],\;\;\;\
\eea
where $c$ is a constant of order unity.  For fixed $T$, $\chi_i^{R,2k_F}(k_\pm,\omega) \to 0$
as $J_0 \to 0$.  This conclusion is independent of the particular form of the operator used in the
spin sector.  For example, using the more general expression \eqref{eq:r_2kF_eff} will lead to
the same conclusion for any ${\cal E}(x,t)$.  Thus, the already weak (because $\left(\frac{J_1}{m\omega_0^2 a^2}\right)^2 \ll 1$) $2k_F$ density oscillations are rapidly suppressed
with temperatures once $T \sim J_0$.  See Fig.~\ref{fig:Wigner_solid} for an illustration.

Having emphasized how ``fragile"  $\chi_i^{R,2k_F}(k_\pm,\omega)$ is for $T\sim J_0$, 
let us now return to the low temperature limit $T \ll J_0$.  In this limit, the temperature dependence of ${\rm Im}\chi_i^{R,2k_F}(k,\omega)=\chi_i^{R,2k_F}(k_+,\omega)+\chi_i^{R,2k_F}(k_-,\omega)$ 
can be readily extracted by making the substitutions $\tilde x =\pi T x/v_c$ and $\tilde t =\pi T t$
and then computing the Fourier transform.  With this substitution, we find
 \be 
 \label{eq:2kFdrag_temp}
 r_D^{2k_F} \propto  \left(\frac{J_1}{m\omega_0^2 a^2}\right)^2 (2k_F)^2  \tilde U^2_{12}(2k_F) T^{2(K_c+K_s)-3}f(T/J_0),
 \ee
 where we have used the result that at low enough temperatures, $\int_0^\infty dk k^2  \tilde U^2_{12}(k) {\rm Im}\chi^R_1(k,\omega) {\rm Im}\chi^R_2(k,\omega)\approx (2k_F)^2  \tilde U^2_{12}(2k_F) {\rm Im}\chi^R_1(2k_F,\omega) {\rm Im}\chi^R_2(2k_F,\omega) \Delta k$, where $\Delta k \sim T$ and used the result that the $\omega$ integration in \eqref{eq:drag_Pustilnik} for \eqref{eq:rho_2kF_temp} 
 does not contribute to any temperature dependence of the drag.  The function $f(X\to0)\to1$
 and $f(X\gtrsim1)\sim e^{-c X}$, where $c$ is a constant of order unity.

 The result \eqref{eq:2kFdrag_temp}
 is identical to the result obtained by Stern and Klesse\cite{Klesse:prb00} in the weakly
 interacting limit of the 1-d electron gas when $T \ll J_0$.  Note that while the temperature dependence is the  same in the low temperature limit, the overall result is still down by a factor $\sim \left(\frac{J_1}{m\omega_0^2 a^2}\right)^2 \ll  1$  when the interactions are strong.

For completeness, it is worth emphasizing that in the high temperature regime ($ T \gg J_0$) the expression \eqref{eq:r_2kF_eff} must be used for
the $2k_F$ part of the density.  In this case, one must compute the Fourier transform of the 
correlator
\bea
\label{eq:rho_2kF_hightemp}
-i\langle [\rho^{\rm eff}_{2k_F}(x,t),\rho^{\rm eff}_{2k_F}(0,0)]\rangle=
\frac{1}{16}\left(\frac{J_1}{m\omega_0^2 a^2}\right)^2\cos(2k_F x) \nonumber \\
\times  {\rm Im}\Biggl[
\frac{(\pi T\alpha/v_c)^{K_c}}{\left[\sinh\left(\frac{\pi T}{v_c}(x- v_ct)\right)\sinh\left(\frac{\pi T}{v_c}(x+ v_ct)\right)\right]^\frac{K_c}{2}}\;\;\;\;
\nonumber \\ 
\times
\langle [{\cal E}(x,t),{\cal E}(0,0)]\rangle\Biggr]. \hspace{2 cm} 
\eea
In the high temperature regime \eqref{eq:rho_2kF_hightemp}  will not behave much differently
from \eqref{eq:rho_2kF_temp} when $T \approx J_0$.  In particular, we expect
\be
\langle [{\cal E}(x,0),{\cal E}(0,0)]\rangle\sim e^{-|x|/\xi_s}\;,
\ee
where $\xi_s \lesssim a$ and so in the high temperature limit the results will be qualitatively
similar to what we discussed earlier.  Of course, the detailed structure of ${\rm Im}\chi_i^{R,2k_F}(k,\omega)$ for
$T \approx J_0$ requires that \eqref{eq:rho_2kF_hightemp} be used.  This in turn requires
that the dimer-dimer correlation function $\langle [{\cal E}(x,t),{\cal E}(0,0)]\rangle$ be
evaluated by a more general (perhaps numerical) method than the effective low energy theory
given in \eqref{eq:S_s}.

\subsubsection{${\rm Im}\chi_i^{R,4k_F}(k,\omega)$}

In the previous subsection we saw that when $v_s/v_c \ll 1$ and temperature 
$T \gg J_0$, the $2k_F$ contributions to the drag
are dramatically suppressed and only the $4k_F$ contributions remain. In contrast to the
case of the $2k_F$ density fluctuations, the present model \eqref{eq:S_c} allows for a closed analytic
expression for ${\rm Im}\chi_i^{R,4k_F}(k,\omega)$.  We begin with the $4k_F$ part of the 
density operator \eqref{eq:rho_eff_final},
\be
 \rho^{\rm eff}_{4k_F}(x,t) = \rho_0 \cos\left(4k_Fx+\sqrt{8}\theta_c(x,t)\right),
 \ee
which leads, after evaluating the correlators at finite temperature, to
\bea
\label{eq:rho_4kF_temp}
-i\langle [\rho^{\rm eff}_{4k_F}(x,t),\rho^{\rm eff}_{4k_F}(0,0)]\rangle=
\rho_0^2 \cos(4k_F x)\hspace{2 cm}\nonumber \\
\times{\rm Im}\Biggl[\frac{(\pi T\alpha/v_c)^{4K_c}}{\left[\sinh\left(\frac{\pi T}{v_c}(x- v_ct)\right)\sinh\left(\frac{\pi T}{v_c}(x+ v_ct)\right)\right]^{2K_c}}\Biggr].\;\;\;\;
\eea
As in the case of ${\rm Im}\chi_i^{R,2k_F}(k,\omega)$ the temperature dependence at low
enough temperatures
can be extracted by making the substitutions $\tilde x =\pi T x/v_c$ and $\tilde t =\pi T t$.
This then leads us to $\chi_i^{R,4k_F}(k,\omega)=\chi_i^{R,4k_F}(k_+,\omega)+\chi_i^{R,4k_F}(k_-,\omega)$
where
\bea
\label{eq:chi_4kF_int}
\chi_i^{R,4k_F}(k_\pm,\omega)=T^{4K_c-2}v_c\int_{-\infty}^\infty \!\!\! d\tilde x  \int_{0}^\infty \!\!\! d\tilde t e^{i\left(\frac{\omega\tilde t -v_ck_\pm \tilde x}{\pi T}\right)}\nonumber \\
\times 
\frac{\rho_0^2}{\pi^2}{\rm Im}\Biggl[ \frac{(\pi \alpha/v_c)^{4K_c}}{\left[\sinh(\tilde x- \tilde t)\sinh(\tilde x+\tilde t)\right]^{2K_c}}\Biggr],\;\;\;\;
 \eea
 and  $k_\pm=k\pm 4k_F$.  By the same arguments made in the previous subsection (that the form
\eqref{eq:chi_4kF_int} substituted into \eqref{eq:drag_Pustilnik} leads to no temperature dependence
of the drag from the $\omega$ integration, and that the dominant contribution from the $k$ integral comes from $k_\pm \approx 0$ with $\Delta k\sim T$) the temperature dependence of the $4k_F$ contribution to the drag is
\be 
\label{eq:4kFdrag_temp}
 r_D^{4k_F} \propto  (4k_F)^2  \tilde U^2_{12}(4k_F) T^{8K_c-3}\;,
\ee
 which is identical to the result \eqref{eq:rhoD_high} obtained in Sec.~\ref{sec:general}
 by applying the general arguments of Ref.~\onlinecite{Fiete_2:prb05} for the mapping of
 a spin incoherent LL to a spinless LL.

Fortunately, the Fourier transform \eqref{eq:chi_4kF_int} can be computed exactly.\cite{Sachdev:prb94,Schulz:prb86} This is done by making the change of variables $s_1=\tilde x -\tilde t$
and $s_2=\tilde x +\tilde t$, and using the integral result\cite{integral}
\be
\int_0^\infty ds\frac{e^{izs}}{[\sinh(s)]^g}=2^{g-1}\frac{\Gamma(g/2-iz/2)\Gamma(1-g)}{\Gamma(1-g/2)},
\ee
to obtain
\bea
\label{eq:xi4kFfinal}
\chi_i^{R,4k_F}(k_\pm,\omega)=-\frac{\rho_0^2\alpha^2}{2\pi v_c}\left(\frac{2\pi T \alpha}{v_c}\right)^{4K_c-2}
\frac{\Gamma(1-2K_c)}{\Gamma(2K_c)}\nonumber \\
\times \frac{\Gamma(K_c-i\frac{\omega+v_ck_\pm}{4\pi T} )\Gamma(K_c-i\frac{\omega-v_ck_\pm}{4\pi T})}{\Gamma(1-K_c-i\frac{\omega+v_ck_\pm}{4\pi T})\Gamma(1-K_c-i\frac{\omega-v_ck_\pm}{4\pi T})}.\hspace{.5 cm}
\eea

\begin{figure}[h]
\includegraphics[width=.8\linewidth,clip=]{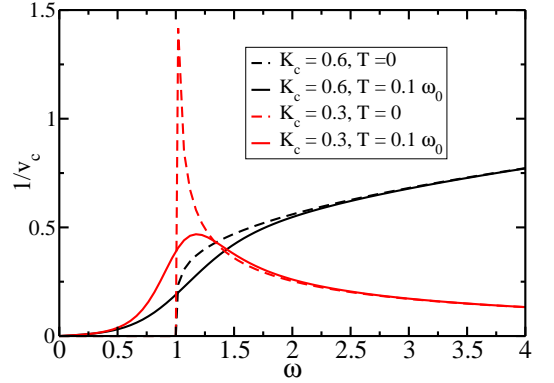}
\caption{\label{fig:xi4kFtemp} (Color online) Frequency dependence of ${\rm Im}\chi_i^{R,4k_F}(k_\pm=1,\omega)$,
Eq.~\eqref{eq:xi4kFfinal}, for various interaction strengths $K_c$ and temperatures $T$.  The
charge velocity is fixed at $v_c=1$.  At $K_c=0.5$ there is a crossover from a sharp peak for $K_c<0.5$
at $\omega=k_\pm$ to monotonically increasing (with $\omega$) threshold-type behavior
for $K_c>0.5$. Finite temperature acts to smear the $T=0$ $\omega=k_\pm$ singularity and adds
weight to ${\rm Im}\chi_i^{R,4k_F}(k_\pm=1,\omega)$ for $\omega < k_\pm$.}
\end{figure}

The temperature dependence of ${\rm Im}\chi_i^{R,4k_F}(k_\pm=1,\omega)$ for different interaction values $K_c$ is shown in Fig.~\ref{fig:xi4kFtemp}.  At $K_c=0.5$ there is a crossover from a divergence to a threshold-type behavior.  The main effect of the temperature is to smooth the sharper features
near $\omega=v_c k$.

\subsection{Weakly Disordered Wires}

\subsubsection{Slowly varying background potential}

Modulation doping in quantum wire systems gives rise to a
smoothly varing background potential.  Such disorder has an important
effect on the drag as it impacts the nature of the electronic states
that participate in drag.\cite{Gornyi:prl05}  Since the coupling of the density to the potential
depends crucially on the Fourier components $k$, there is an 
important difference in how the
charge density couples to disorder in the spin coherent and spin
incoherent regimes. Consider the following coupling of the density to
background potential modulations:
\bea
H_{\rm dis}=\int dx V(x)\rho(x) \hspace{5 cm}\nonumber \\
\approx \int dx \biggl[ V_{0}(x)\rho_{0}(x)
+V_{2k_F}(x)\rho_{2k_F}(x)+V_{4k_F}(x)\rho_{4k_F}(x)\biggr].\nonumber \\
\eea
Here $V_{2k_F}(x)={\rm Re}[V_{2k_F} e^{i2k_F x }]$ and $V_{4k_F}(x)={\rm Re}[V_{4k_F}
e^{i4k_Fx}]$. We can study the scaling dimensions of $V_{2k_F},V_{4k_F}$ 
using the expression for the density, Eq.~\eqref{eq:rho_eff_final},
after integrating out the high energy field $u_\pi$ in favor of the lower 
energy spin
fields.  The scaling dimensions of the different scattering terms can then be 
determined from the action (where the integration over $x$ has already been caried out)
\be 
S_{\rm dis}^{2k_F}\sim \int d\tau V_{2k_F} e^{i\sqrt{2}\theta_c}  e^{i\sqrt{2}\theta_s}+h.c.\;,
\ee
and
\be 
S_{\rm dis}^{4k_F}\sim \int d\tau V_{4k_F} e^{i\sqrt{8}\theta_c}+h.c.\;,
\ee
which gives
\be
{\rm dim}[V_{2k_F}]=1-\frac{K_c}{2}-\frac{K_s}{2},
\ee
\be 
{\rm dim}[V_{4k_F}]=1-2K_c.
\ee
In these units $SU(2)$ invariance implies $K_s=1$, so that the
$2k_F$ piece is more relevant than the $4k_F$ piece of the potential
whenever $K_c>1/3$.  Thus, we expect to see strong temperature
dependence of the pinning of the density whenever $K_c>1/3$ as the
more relevant $2k_F$ piece will be lost for $T \gg J_0$.   Moreover, if $1/2 < K_c <1$ the
$2k_F$ piece is relevant while the $4k_F$ piece is irrelevant.  In
this case, the effect should be most dramatic.  The regime $1/2 < K_c <1$
can be reached for large but finite $U$ in a one dimensional Hubbard model.

\subsubsection{Effects of random correlations in forward scattering on the correlation functions}

As the $k\approx 0$ part of the back ground potential fluctuations are often the
most important at low energies, it is worthile to
review\cite{Voit:rpp95} their influence on the density-density
correlation function.  The forward scattering part of the background
potential is
\be
H_{\rm dis}^f=\int dx V_{0}(x)\rho_{0}(x)=-\frac{\sqrt{2}}{\pi} \int dx V_{0}(x)\partial_x\theta_c(x)\;,
\ee
where we have we have used the 
result \eqref{eq:rho_eff_final} and assumed $\int dx V_{0}(x) \rho_0=0$.  Forward scattering can be completely
eliminated from the Hamiltonian (action) by making the change of variables
\be
\tilde \theta_c(x)=\theta_c(x)-\frac{\sqrt{2}K_c}{v_c}\int^x \!\!\!dz V_0(z)\;,
\ee
and completing the square in Eq.\eqref{eq:S_c}.  Assuming that the disorder
has white noise correlations given by the distribution 
$P_{V_0}={\rm exp}[-D^{-1}\int dz |V_0(z)|^2]$,
\be 
\label{eq:white_noise}
\overline{ V_0(x)V_0(x')}=\frac{D}{2}\delta(x-x')\;,
\ee
where the overbar indicates a disorder average, and the 
constant $D\sim v_c/\tau_{\rm scatt}$ with $\tau_{\rm scatt}$
the typical scattering time for the electrons.  The disorder averaged
parts of the density-density correlation function can then readily be
determined:
\be
\overline{\langle [\rho_0^{\rm eff}(x,t),\rho_0^{\rm eff}(0,0)]\rangle}
=\langle [\rho_0^{\rm eff}(x,t),\rho_0^{\rm eff}(0,0)]\rangle|_{V_0=0},
\ee
\bea
\overline{\langle [\rho_{2k_F}^{\rm eff}(x,t),\rho_{2k_F}^{\rm eff}(0,0)]\rangle}=\hspace{3 cm}
\nonumber \\
e^{-D\left(K_c \over v_c\right)^2|x|}\langle [\rho_{2k_F}^{\rm eff}(x,t),\rho_{2k_F}^{\rm eff}(0,0)]\rangle|_{V_0=0},
\eea
\bea
\overline{\langle [\rho_{4k_F}^{\rm eff}(x,t),\rho_{4k_F}^{\rm eff}(0,0)]\rangle}=\hspace{3 cm}
\nonumber \\
e^{-4D\left(K_c \over v_c\right)^2|x|}\langle [\rho_{4k_F}^{\rm eff}(x,t),\rho_{4k_F}^{\rm eff}(0,0)]\rangle|_{V_0=0}.
\eea
 It is evident 
that larger wavevectors are suppressed more by the forward scattering with no suppression
at all for the $k\approx 0$ part of the density.  
Treating the $2k_F$ and $4k_F$ backscattering contributions
with white noise correlations analogous to \eqref{eq:white_noise} is more
involved and requires studying renormalization group 
flows.\cite{Giamarchi:prb88} However, we reiterate that if the disorder
is slowly-varying (as expected from donor potential modulations in
modulation doped structures), the $2k_F$ and $4k_F$ potentials are
relatively weak and probably negligible, at least in the sorts of
structures optimized for the drag measurements we envision here!

\subsubsection{Fourier transform of disorder averaged correlation functions}

Once the Fourier transforms of the $2k_F$ and $4k_F$ density-density correlation functions
\eqref{eq:rho_2kF_temp} and \eqref{eq:rho_4kF_temp} are known, the Fourier transforms of
the disorder averaged correlation functions are readily computed from the convolution theorem
\be
\label{eq:disorderedxi4kF}
\overline{{\rm Im}\chi^R_i(k,\omega)}=\int_{-\infty}^{\infty}\frac{dq}{2\pi} \frac{2l^{-1}}{l^{-2}+(k-q)^2} {\rm Im}\chi^R_i(q,\omega)\;,
\ee
where $l^{-1}=D(K_c/v_c)^2$ for the $2k_F$ pieces and $l^{-1}=4D(K_c/v_c)^2$ for the $4k_F$ pieces.
The main effect of the disorder is thus to broaden the singularities in $\chi^R_i(k,\omega)$ by
an amount of order $l^{-1}$ in momentum space.  Hence, the $4k_F$ singularities are broadened
4 times as much as the $2k_F$ singularities.

\begin{figure}[h]
\includegraphics[width=.8\linewidth,clip=]{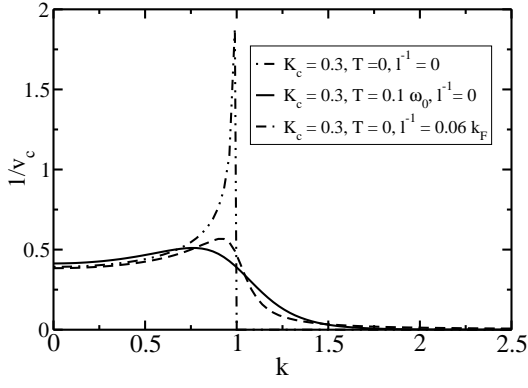}
\caption{\label{fig:xi4kFdisorder} Disorder and thermal broadening of ${\rm Im}\chi^{R,4k_F}_i(k,\omega=1)$.  The charge velocity is fixed at $v_c=1$.  The disorder broadening is computed using Eq.~\eqref{eq:disorderedxi4kF}. The effects of weak forward scattering on the $4k_F$ density fluctuations
are qualitatively similar to that of thermal broadening--see case of $l=0.06 \,k_F$.  Analagous results are also obtained for
the $2k_F$ density fluctuations.  }
\end{figure}

Fig.~\ref{fig:xi4kFdisorder} illustrates the effect of weak disorder on ${\rm Im}\chi^{R,4k_F}_i(k,\omega)$.
Qualitatively the effects of disorder are very similar to finite temperature--there is a broadening of the
sharpest features in the response.  This has implications for the temperature dependence of the
drag as we discuss in the next section.

\section{Study of the Coulomb Drag}
\label{sec:drag}

We have already discussed several features of the drag in the previous sections of this paper,
including the temperature dependence (in certain limits) of the $k\approx 2k_F$ and $k\approx 4k_F$ contributions to the drag.  Implicit in those discussions was that the wires were identical.  In this
section we will present numerical calculations of the Coulomb drag as a function of temperature for
identical and non-identical wires, and attempt to illuminate the crossover to the spin incoherent
regime with semi-quantitative estimates.

\subsection{Drag at low temperatures and in the crossover to the spin incoherent regime}

\subsubsection{The low temperature Luttinger liquid regime}

At the lowest temperatures where $T \ll J_0$ we showed that the low energy theory \eqref{eq:S_c},
\eqref{eq:S_s} results in the following temperature dependence of the drag resistivity
\eqref{eq:drag_Pustilnik}
\bea
\label{eq:drag_lowtemp}
r_D \approx  A(K_c,v_c)(4k_F)^2  \tilde U^2_{12}(4k_F) \left(\frac{T}{T_{4k_F}}\right)^{8K_c-3}\hspace{1 cm}\nonumber \\
+B(K_c,K_s,v_s/v_c)(2k_F)^2 \tilde U^2_{12}(2k_F) \left(\frac{T}{T_{2k_F}}\right)^{2(K_c+K_s)-3}
\eea
where $A(K_c,v_c)$ and $B(K_c,K_s,v_s/v_c)$ are functions that depend on the variables
indicated and $T_{2k_F}^{-1}=\pi\alpha/\sqrt{v_cv_s}$ and $T_{4k_F}^{-1}=\pi\alpha/v_c$ are effective
temperatures in the respective sectors.    It is
evident from \eqref{eq:drag_lowtemp} that the temperature dependence of the $k\approx 2k_F$ and $k\approx 4k_F$ contributions are different so there is a temperature at which the two balance out:
\be
\label{eq:Tstarstar}
\frac{T^{**}}{T_{2k_F}}=\left[\left(\frac{T_{4k_F}}{T_{2k_F}}\right)^{8K_c-3}\frac{B(K_c,K_s,v_s/v_c) \tilde U^2_{12}(2k_F)}{4 A(K_c,v_c)\tilde U^2_{12}(4k_F)}
\right]^{\frac{1}{6K_c-2K_s}}\hspace{-1 cm},
\ee
From here on, we will assume $SU(2)$ symmetry which implies $K_s=1$.  When $K_s=1$ it
is clear that the $2k_F$ contribution to the drag is an increasing function of temperature whenever
$K_c>1/2$ and a decreasing function otherwise.  For the $4k_F$ contribution the boundary between
increasing and decreasing contributions is $K_c=3/8$.   Finally, by comparing the exponents of the
$2k_F$ and $4k_F$ terms, one finds that the $4k_F$ pieces dominate the drag for $K_c>1/3$ when $T>T^{**}$, while the $2k_F$ pieces dominate the drag for  $K_c<1/3$ in the same temperature
regime.  Note that this implies that the $2k_F$ density oscillations are more important for the drag
at {\em higher} temperatures when the interactions are strong enough that $K_c<1/3$.  This requires,
of course, that the system is still at low enough temperatures that the spin degrees of freedom can
be described by the effective low energy theory \eqref{eq:S_s}.  In order to obtain $T^{**} < J_0$ and
for the analysis above to be reasonable, we expect that we must have $\tilde U(2k_F) \gg \tilde U(4k_F)$.

From the results of Appendix~\ref{app:AB} we can express the ratio
\be
\label{eq:AB_ratio}
\frac{B(K_c,K_s,v_s/v_c)}{A(K_c,v_c)}=\left(\frac{J_1}{16\pi m\omega_0^2a^2}\right)^4
\frac{I_{2k_F}(K_c,K_s,v_s/v_c)}{I_{4k_F}(K_c)},
\ee
where $I_{2k_F}(K_c,K_s,v_s/v_c)$ and $I_{4k_F}(K_c)$ are given by  Eqs.~\eqref{eq:I_2kF}
and \eqref{eq:I_4kF}.   As $v_s/v_c \to 0$ the crossover temperature \eqref{eq:Tstarstar} becomes
very small because both $I_{2k_F}(K_c,K_s,v_s/v_c) \ll I_{4k_F}(K_c)$ and $\left(\frac{J_1}{16\pi m\omega_0^2a^2}\right)^4\sim \left(\frac{v_s}{v_c}\right)^4 \ll 1$ in that limit.  Of course, as $v_s$
shrinks for fixed $v_c$, the temperature range over which the LL theory itself is valid is also shrinking
and the spin incoherent regime (where only the $4k_F$ density modulations remain) is approached.

\subsubsection{Crossover to the spin incoherent regime}

The hallmark of the spin incoherent regime is the equivalence of the real electron system with spin to a spinless system\cite{Fiete_2:prb05} with the exception of the Green's function\cite{Fiete:prl04} and
other non-particle conserving operators.  In the case of drag, spin incoherence manifests itself as a
thermal washing out of the $2k_F$ oscillations in the density-density correlation function \eqref{eq:chi}.
When the interactions are as strong as they are here, the weight of the $2k_F$ oscillations are already
down by  a factor $\sim\left(\frac{J_1}{16\pi m\omega_0^2a^2}\right)^2$ even at zero temperature.

As we have discussed before,\cite{Fiete:prl04} the spin incoherent regime can be understood by starting with $T\gg J_0$, taking $J_0\to 0$, for fixed $T$ and then finally taking $T\to0$.  In the
present formulation this is equivalent to fixing a finite, but low temperature, applying the low energy
theory \eqref{eq:S_c} and \eqref{eq:S_s} and then taking the limit $v_s/v_c \to 0$ as we did in the
previous section.  The approach to the spin incoherent drag regime can be directly obtained via this procedure.  One expects that as $v_s$ is lowered, the power law
\eqref{eq:2kFdrag_temp} will first breakdown (at temperatures it once held for larger $v_s$) before
the contribution vanishes altogether from $f(T/J_0)$.

\subsection{Drag in the spin incoherent regime}

In this subsection we present some numerical results justifying earlier analytical arguments for
the temperature dependence of the drag.  We first consider identical wires and then we study
non-identical wires.

\subsubsection{Identical Wires}

\begin{figure}[h]
\includegraphics[width=.8\linewidth,clip=]{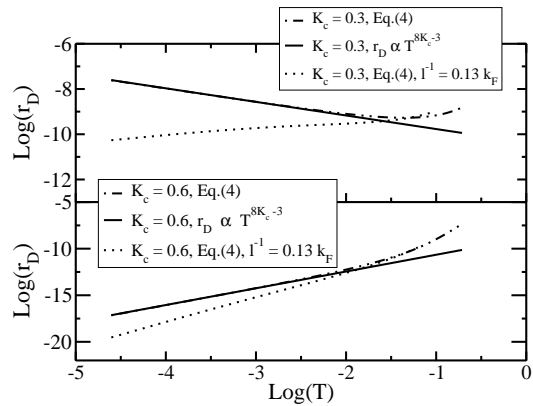}
\caption{\label{fig:dragT} Temperature dependence of the Coulomb drag in the spin
incoherent regime where only the $4k_F$ components of the density fluctuations contribute.  The drag
was computed over a temperature range $0.01 \omega_0 \leq T \leq 0.5 \omega_0$. Shown
is a comparison between the drag formula  \eqref{eq:drag_Pustilnik}  with  \eqref{eq:xi4kFfinal} substituted in and the approximate power law \eqref{eq:4kFdrag_temp}.  Two values of the interaction parameter $K_c$ are shown.  A disorder value of $l^{-1}=0.13 k_F$ was used in Eq.~\eqref{eq:disorderedxi4kF}
to compute the drag from Eq.~\eqref{eq:drag_Pustilnik}.  For $k_BT \gg \hbar v_c /l$ the  drag for the clean and disordered system are indistinguishable, while for $k_B T \ll \hbar v_c /l$
there is a crossover of the temperature dependence to another power of the temperature.
Empirically, we found the power law $r_D^{4k_F,\;\rm disorder} \propto T^{8K_c-2}$ to be a very good fit for
any value of $0<K_c<1$. }
\end{figure}

When the wires are identical we expect the temperature dependence of the Coulomb drag given
in Eq.~\eqref{eq:drag_lowtemp} to be obtained at the lowest energies.  However, as we have seen
in the previous sections the $2k_F$ oscillations are rapidly washed out in the limit $v_s/v_c \to 0$
and only the $4k_F$ oscillations remain.  In this subsection, we provide numerical evidence that
the manipulations leading to the $4k_F$ temperature dependence of  \eqref{eq:drag_lowtemp} is justified.  Since these are also the same arguments leading to the $2k_F$ temperature dependence
of $r_D$ at the lowest temperature, these are implicitly justified as well.  Fig.~\ref{fig:dragT} illustrates the comparison between the exact result from \eqref{eq:xi4kFfinal} substituted into \eqref{eq:drag_Pustilnik}, and the approximate power law \eqref{eq:4kFdrag_temp}.  A disorder value of $l^{-1}=0.13 k_F$ was used in Eq.~\eqref{eq:disorderedxi4kF} to compute the drag of the disordered system from Eq.~\eqref{eq:drag_Pustilnik}. The drag was computed over a temperature range $0.01 \omega_0 \leq T \leq 0.5 
\omega_0$.  For $k_BT \gg \hbar v_c/ l$ the  drag of the clean and disordered system are indistinguishable, while for $k_BT \ll \hbar v_c/l$ there is a crossover of the temperature dependence to another power of the temperature.
Empirically, we found the power law 
\be
r_D^{4k_F, \rm disorder} \propto T^{8K_c-2}
\ee
to be a very good fit for
any value of $0<K_c<1$.  This temperature dependence can actually be inferred from  \eqref{eq:chi_4kF_int}, \eqref{eq:drag_Pustilnik} and Fig.~\ref{fig:xi4kFtemp}.  As we have argued several times earlier, the $\omega$ integration in \eqref{eq:drag_Pustilnik}  does not contribute any temperature dependence beyond the $(T^{4K_c-2})^2$ factors in front of   $\eqref{eq:chi_4kF_int}$ (with the square coming from the drag formula  \eqref{eq:drag_Pustilnik}).  When $k_BT \ll \hbar v_c /l$, the $k$ integration picks up a contribution proportional to $T^2$ rather than $T$.  Adding
up the exponents leads to $r_D^{4k_F, \rm disorder} \propto T^{8K_c-2}$.

\subsubsection{Drag for non-identical wires}

\begin{figure}[h]
\includegraphics[width=.8\linewidth,clip=]{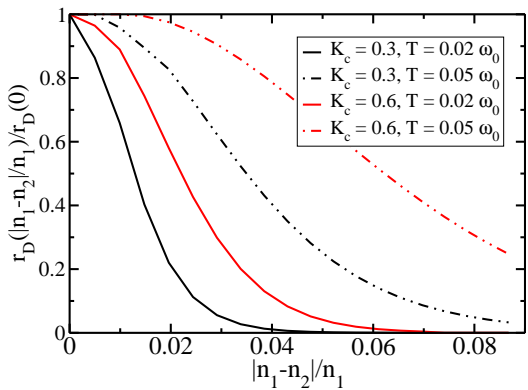}
\caption{\label{fig:diff_dens} (Color online) Coulomb drag dependence on relative wire density at fixed temperature. We have set $v_c=1$. The Coulomb drag between two wires drops rapidly as a function of density mismatch.
Only a few percent change in the relative densities of the wires results in a dramatic suppression of the drag.  At higher temperatures and weaker interactions the drag response is more robust to small density differences between the wires. }
\end{figure}

Coulomb drag for non-identical wires and the incommensurate-commensurate transition has been discussed for fully coherent clean wires in Ref.~\onlinecite{Fuchs:prb05}.  Via the mapping detailed in Ref.~\onlinecite{Fiete_2:prb05} the incommensurate-commensurate transition can be ready discussed
deep in the spin incoherent regime.  In Fig.~\ref{fig:diff_dens} we present some numerical results for
the dependence of the drag for small density mismatches between the two wires.  Note that weaker interactions and higher temperatures lead to a more robust drag effect between two wires of slightly different densities.  Note also that with only a few percent change in the relative densities of the wires the drag effect is substantially reduced.  

\begin{figure}[h]
\includegraphics[width=.8\linewidth,clip=]{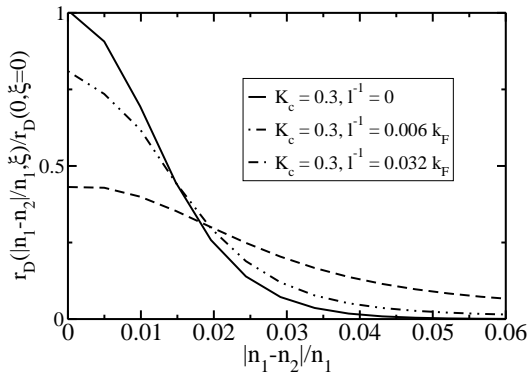}
\caption{\label{fig:diff_dens_xi} Coulomb drag dependence on relative wire density at fixed temperature for different values of disorder. Here $T = 0.02\, \omega_0$ and $v_c=1$.  While for clean wires the Coulomb drag drops rapidly as a function of density mismatch, some amount of residual disorder allows for a more robust drag effect between quantum wires that may have slightly different values of $k_F$.}
\end{figure}

The effects of disorder on the drag for density mismatched wires is shown in Fig.~\ref{fig:diff_dens_xi}.
When $\hbar v_c/ l \gtrsim k_B T$ the disorder has a significant effect--making the drag more robust
for non-identical wires.  While for small $|n_1-n_2|$ the drag is reduced relative to the clean limit, for larger $|n_1-n_2|$ there can be appreciable enhancement.

Finally, we note that in the case of clean wires of different charge velocities but the same Fermi wavevector (in which case the $K_c$ are diffferent) the temperature dependence of the drag is
\be
r_D^{4k_F} \sim T^{4(K_{c,1}+K_{c,2})-3}\;,
\ee
and if the temperature is also much less than $J_0$,
\be
r_D^{2k_F} \sim T^{K_{c,1}+K_{c,2}+K_{s,1}+K_{s,2}-3}\;.
\ee

\section{Summary of Results}
\label{sec:summary}

We have discussed the Coulomb drag between two quantum wires in the
limit of low electron density where at finite temperatures the energy
hierarchy $J \ll T \ll E_F$ can be obtained.\cite{Steinberg:prl06}  In this limit, the spin
degrees of freedom are completely incoherent and we have shown this
implies the loss of $2k_F$ oscillations in the density-density
correlation function.  As a result, a non-monotonic temperature dependence of the drag
on temperature may result. In the spin incoherent Luttinger liquid regime,
the drag problem maps onto the identical problem for a spinless
Luttinger liquid only with $K_{c-}\to 2 K_{c-}$ so that for a clean
wire the drag resistivity goes as $\rho_D \propto T^{8K_{c-}-3}$,
where $K_{c-}$ is the coupling parameter of the antisymmetric charge
mode of a Luttinger liquid theory with spin.  We have shown this temperature dependence explicitly
with approximate analytical calculations and confirmed those approximations with numerics.

Our results are based on a fluctuating Wigner solid model appropriate to quantum wires in a very strongly interacting regime, which typically implies low electron density.  The spin sector is modeled
as a nearest-neighbor antiferromagnetic Heisenberg spin chain.  Without any coupling of the spin
and charge sectors, no $2k_F$ density modulations appear.  However, including a magneto-elastic coupling term that allows for a linear change in the nearest-neighbor spin coupling for small distortions induces $2k_F$ oscillations in the density.  The coupling is weak,  however, and the $2k_F$ oscillations are easily washed out by temperatures $T\gtrsim J$.

The fluctuating Wigner solid model is studied by deriving effective expressions for the density operator when the highest energy phonon modes are integrated out in favor of spin operators.   At the lowest energies (including the spin energy) an expression is obtained equivalent to that known well in Luttinger liquid theory except the $2k_F$ terms contain a prefactor of order the ratio of kinetic to potential energy.  Nevertheless, in this limit all correlations functions exhibit power law decay with the familiar exponents of the spin and charge sectors.   

These density operators are then used to compute density-density correlation functions which are Fourier transformed into frequency and momentum space and used in previously derived drag formulas.  Since the density operator has contributions at momenta $k \approx 0,\;k\approx 2k_F$, and $k \approx 4k_F$
the Coulomb drag will generically contain contributions from each of these pieces.  We show explicitly that the $k\approx 0$ piece vanishes due to the harmonic approximation to the Wigner solid.  This is equivalent to linearizing the electron dispersion in the standard Luttinger liquid treatment for weakly interacting electrons. Generically, the $k\approx 2k_F$ and $k \approx 4k_F$ contributions are non-vanishing and we explicitly compute their temperature dependence, $T^{2K_c-1}e^{-cT/J_0}$ and  $T^{8K_c-3}$, in the low temperature regimes.

We have also considered the case of non-ideal wires in which weak disorder is present.  We find
that white-noise correlated forward scattering disorder does not affect the  $k\approx 0$ result, 
while it tends to broaden the sharp $k-$space features of the $k\approx 2k_F$ and $k \approx 4k_F$
density-density correlation function in a manner similar to temperature.  As a result, the drag resistivity 
crosses over to a different power law, $T^{2K_c} $ and $T^{8K_c-2}$, which is increase by one power of the temperature relative to the clean result.  Finally, we have also studied the reduction of the drag due to a density mismatch between the two wires and shown the drag may be substantially reduced with only a few percent change in the relative densities of the wires.  When disorder is present the drag is more robust to density mismatches between the two wires and this fact is likely to play an important role in real drag experiments. 

We hope that this work will help to inspire further experimental studies on one-dimensional drag, which to date is quite limited.

\acknowledgments

This work was supported by NSF Grant numbers
PHY99-07949, DMR04-57440, the Packard Foundation, CIAR, FQRNT, and
NSERC.

\appendix
\section{Exact expressions for $\langle \rho(x,\tau)\rho(x',\tau')\rangle$ up to second order in $J_1$}
\label{app:dens_exact}

The low-energy description given in Sec.~\ref{sec:Wigner} can be treated more accurately, but in a less physically transparent way by applying the results of this appendix.

\subsection{Diagonalization of $H_c$}

The charge Hamiltonian (\ref{eq:H_c}) is diagonalized by the transformation
\begin{eqnarray}
u_l&=&\frac{1}{\sqrt N} \sum_k e^{ikal}u_k\;,\\
p_l&=&\frac{1}{\sqrt N} \sum_k e^{-ikal}p_k\;,
\end{eqnarray}
(assuming periodic boundary conditions) where the theory is quantized by imposing $[u_l,p_m]=i\hbar \delta_{lm}$ and $[u_k,p_{k'}]=i\hbar \delta_{k k'}$.  Making these substitutions we find
\begin{equation}
\label{eq:H_c_k}
H_c=\sum_{k=k_1}^{k_N} \frac{p_kp_{-k}}{2m} +\frac{m\omega_k^2}{2}  u_k u_{-k}\;,
\end{equation}
where $\omega_k=2 \omega_0 | \sin\left(\frac{ka}{2}\right)| \approx v_c |k|$ for small $k$ with
$v_c=\omega_0 a$ the sound velocity of the charge modes.  In momentum space 
the harmonic chain is just a sum of harmonic oscillators with frequencies that depend on the wavenumber $k$.

The Hamiltonian (\ref{eq:H_c_k}) can be brought into a particularly simple form via the transformation
\begin{eqnarray}
a_k &=& \sqrt{\frac{m\omega_k}{2\hbar}}\left(u_k + \frac{i}{m\omega_k}p_{-k}\right)\;,\\
a^\dagger_k &=& \sqrt{\frac{m\omega_k}{2\hbar}}\left(u_{-k} - \frac{i}{m\omega_k}p_{k}\right)\;,
\end{eqnarray}
which brings $H_c$ to
\begin{equation}
\label{eq:H_c_a}
H_c=\sum_k \hbar \omega_k \left(n_k+\frac{1}{2}\right)\;.
\end{equation}
For later reference it is useful to note that 
\begin{equation}
\label{eq:u_l}
u_l(t)=\sum_k  M_k(l) \left(a_ke^{-\omega_k \tau} + a^\dagger_{-k} e^{\omega_k \tau}\right)\;,
\end{equation}
where $M_k(l)\equiv \sqrt{\frac{\hbar}{2N m\omega_k}} e^{ikal}$ and we have used $\omega_k=\omega_{-k}$.

\subsection{Notation for perturbation theory}

The general expression for the density-density correlation function 
at finite temperature $T=1/\beta$ is
\begin{equation}
\label{eq:rho_rho_gen}
\langle \rho(x,\tau)\rho(x',\tau')\rangle = \frac{\langle U(\beta)\rho(x,\tau)\rho(x',\tau')\rangle_0}{\langle U(\beta)\rangle_0}\;,
\end{equation}
where
\begin{equation}
U(\beta)=\sum_{n=0}^\infty (-1)^n \int_0^\beta d\tau_1... \int_0^{\tau_{n-1} }\hat H'(\tau_1)...\hat H'(\tau_n)\;,
\end{equation}
with the operators taken in the interaction representation. The averages $\langle ... \rangle_0\equiv {\rm Tr}[e^{-\beta H_0}...]$ where the $0^{th}$ order Hamiltonian
is (\ref{eq:H_wire}) taken with $J_1\equiv 0$:
\begin{equation}
H_0=H_c + J_0\sum_l {\vec S_{l+1}}\cdot{\vec S_l} \;,
\end{equation}
and $H'$ is the correction to this to be treated in perturbation theory
\begin{equation}
\label{eq:H'}
H'= J_1\sum_l (u_{l+1}-u_l){\vec S_{l+1}}\cdot{\vec S_l} \;.
\end{equation}

\subsection{Evaluation of the density-density correlation function}

\subsubsection{Zeroth Order}
At zeroth order we have $J_1\equiv 0$ and there is no coupling between
the charge and the spin degrees of freedom.  Therefore we can completely
ingore the spin sector since it traces out trivially.  Thus,
\begin{equation}
\langle \rho(x,\tau)\rho(x',\tau')\rangle^{(0)} = Z_c^{-1} {\rm Tr}[e^{-\beta H_c}\rho(x,\tau)\rho(x',\tau')]\;,
\end{equation}
where $Z_c={\rm Tr}[e^{-\beta H_c}]$ is the partition function of 
the charge sector and the density is expressed as $\rho(x,\tau)=\sum_{l=1}^N \delta(x-al-u_l(\tau))$.  From the Hamiltonian (\ref{eq:H_c_a}) $Z_c$ can be readily
evaluated as 
\begin{equation}
Z_c=\prod_k \sum_{n_k=0}^\infty e^{-\beta \hbar \omega_k (n_k+1/2)}=\prod_k e^{-\beta \hbar \omega_k/2}\frac{1}{1-e^{-\beta\hbar \omega_k}}\;,
\end{equation}
which we will make use of later.  Thus,
\bea
\langle \rho(x,\tau)\rho(x',\tau')\rangle^{(0)}=Z_c^{-1} \sum_{l,l'}\int\frac{d\eta}{2\pi}\frac{d\xi}{2\pi} e^{i\eta(x-al)} e^{i\xi(x'-al')}\nonumber \\ 
\times {\rm Tr}\left[ e^{-\beta H_c} e^{-i\eta u_l(\tau)} e^{-i\xi u_{l'}(\tau')}\right],\hspace{1 cm}
\eea
where the quantities $u_l$ appearing in the exponent of the trace can be expressed using Eq.~(\ref{eq:u_l}). Using $e^{A+B}=e^A e^B e^{-[A,B]/2}$ 
(where $[A,B]$ commutes with both $A$ and $B$ separately), we focus on the 
trace and obtain
 \bea
&&{\rm Tr}[ e^{-\beta H_c} e^{-i\eta u_l(\tau)} e^{-i\xi u_{l'}(\tau')}]=e^{-\frac{1}{2}\sum_q |M_q|^2(\eta^2+\xi^2)} \nonumber\\
&\times& e^{-\eta \xi \sum_q |M_q|^2 e^{iqa(l-l')-\omega_q(\tau-\tau')}}
\nonumber \\
&\times& 
\prod_k {\rm Tr}^{(k)}\left[e^{-\beta \hbar \omega_k(a^\dagger_ka_k+1/2)}e^{-iC^*(k)a^\dagger_k}e^{-iC(k)a_k}\right],\nonumber \\
\eea
where $C(k)\equiv |M_k|(\eta e^{ikal-\omega_k\tau)}+ \xi
e^{ikal'-\omega_k \tau')})$.  Evaluating the trace for each $k$ independently
and using definition of a Laguerre polynomial of order $n_k$,  
\be
L_{n_k}(-|C(k)|^2)\equiv  \langle n_k| e^{-iC^*(k)a^\dagger_k}e^{-iC(k)a_k}|n_k\rangle\;,
\ee
and then applying the important formula
\be
\sum_{n=0}^\infty L_{n}(|C|^2) z^{n}=\frac{1}{1-z}{\rm exp}\left\{|C|^2\frac{z}{z-1}\right\}\;,
\ee
we find
\bea
\langle \rho(x,\tau)\rho(x',\tau')\rangle^{(0)}=\sum_{l,l'}\int\frac{d\eta}{2\pi}\frac{d\xi}{2\pi} e^{i\eta(x-al)} e^{i\xi(x'-al')}\nonumber \\ 
\times e^{-\frac{1}{2}F(\eta^2+\xi^2)} e^{-\eta \xi G(l,l')}\;,\hspace{.6 cm}
\eea
where
\be
\label{eq:F}
F\equiv \sum_q |M_q|^2 \biggl[1  + \frac{2 e^{-\beta \hbar \omega_q}}{1-e^{-\beta \hbar \omega_q}} \biggr]\;,
\ee
and 
\be
\label{eq:G}
G(l,l')\equiv\sum_q |M_q|^2 \biggl[e^{i\theta_q(l,l';t,t')} +\frac{2\cos\left(\theta_q(l,l';t,t')\right)e^{-\beta \hbar \omega_q}}{1-e^{-\beta \hbar \omega_q}}\biggr]\;,
\ee 
and $\theta_q(l,l';t,t')=qa(l-l')-\omega_q(t-t')$, where $t=-i\tau$.
Finally, shifting the summation variables $\tilde l = l-l'$ and performing
the integrations we obtain
\bea
\label{eq:rho_0_final}
&\langle \rho(x,t)\rho(x',t')\rangle^{(0)}& \nonumber \\
&=& \hspace{-1.4 cm} \frac{1}{a}\sum_{l} \frac{1}{2\pi}\sqrt{\frac{\pi}{F-G(l)}} {\rm exp}\left\{-\frac{(x-x'-al)^2}{4(F-G(l))}\right\},\nonumber \\
\eea
where
\bea
\label{eq:F-G}
F-G(l)=\frac{a\hbar}{mv_c2\pi} \Biggl[ \ln\biggl[\frac{\alpha^2+(al)^2+(v_c(\tau-\tau'))^2}{\alpha^2}\biggr]\nonumber \\
+ 2 \sum_{n=0}^\infty \ln\biggl[\frac{(\alpha +(n+1)\beta \hbar v_c)^2 +(al)^2+(v_c(\tau-\tau'))^2}{(\alpha +(n+1)\beta \hbar v_c)^2}\biggr]\Biggr].\nonumber \\
\eea
Here $\alpha$ is a short distance cut off of order the lattice spacing
$a$.  Note that for finite temperatures the second sum is cut off when
the argument of the $\ln$ becomes ${\cal O}(1)$ which occurs when
$n=n_l\sim \sqrt{(al)^2+(v_c(\tau-\tau'))^2}/(\beta \hbar v_c)$.  In the limit of
$T\to0$, $\beta \to \infty$ and the second terms drops out all
together. In this paper we are interested in the limit $T\ll E_F$, so
the second term can be ignored altogether.  We will not explicitly
consider finite temperatures in the first and second order
expressions.

\subsubsection{First order}

The manipulations needed here are identical to those used to compute
the zeroth order result, so we simply quote the result:
\bea
\langle \rho(x,\tau)\rho(x',\tau')\rangle^{(1)}= J_1 N \sum_{n} \int_0^\beta d\tilde\tau \langle 
{\vec S_{l+1}}\cdot{\vec S_l}(\tilde \tau)\rangle^{(0)} \nonumber \\
\times \frac{I(n)}{2\pi a} \prod_{k>0}^{k_{\rm max}}|M(k)|^4 4(1-\cos(ka))
\nonumber \\
\times (1-\cos\left(\theta_k(n,0;\tau,\tau')\right))e^{-2\tilde\tau\hbar\omega_k},\nonumber \\
\eea
where $l$ is arbitrary, $N$ is the number of electrons in the system,
and 
\bea
I(n)=e^{\frac{B^2}{4A}}\sum_{j=0}^N\frac{\sqrt{\pi}(-1)^j(2N)!}{(2j)!(2N-2j)!}\left(\frac{B}{2A}\right)^{2(N-j)}\nonumber \\
\times \frac{(2j-1)!!}{2^j}A^{-\frac{1}{2}-j}\;,
\eea
where the $n$ dependence enters through $B\equiv i(x-x'-an)$ 
and $A\equiv F-G(n)$.  It
is worth noting that neither $\langle \rho(x,\tau)\rho(x',\tau')\rangle^{(0)}$
nor $\langle \rho(x,\tau)\rho(x',\tau')\rangle^{(1)}$ contain a $2k_F$ component.
This component will only appear in the second order term, as we now discuss.

\subsubsection{Second order}

The second order corrections are (where ${\bf x}=(x,\tau)$)
\bea
\label{eq:rho_rho_2}
 \langle \rho({\bf x})\rho({\bf x'})\rangle^{(2)} = \int_0^{\beta} d\tau_1 \int_0^{\tau_1} d\tau_2  \frac{\langle\hat H'(\tau_1)\hat H'(\tau_2)\rho({\bf x})\rho({\bf x'})\rangle_0}{Z_c Z_s}\nonumber \\
-\langle \rho({\bf x})\rho({\bf x'})\rangle^{(0)}  \int_0^\beta d\tau_1 \int_0^{\tau_1} d\tau_2  \frac{\langle\hat H'(\tau_1)\hat H'(\tau_2)\rangle_0}{Z_c Z_s}\nonumber \\
=(J_1)^2\sum_{l,l'}\int_0^{\beta} d\tau_1 \int_0^{\tau_1}d\tau_2 \langle {\vec S_{l+1}}\cdot{\vec S_l}(\tau_1) {\vec S_{l'+1}}\cdot{\vec S_{l'}}(\tau_2)\rangle^{(0)}\nonumber \\
\times \biggl[ \frac{\langle\hat H'_c(l,\tau_1)\hat H'_c(l',\tau_2)\rho({\bf x})\rho({\bf x'})\rangle_0}{Z_c}\nonumber \\
-\langle \rho({\bf x})\rho({\bf x'})\rangle^{(0)} \frac{\langle\hat H'_c(l,\tau_1)\hat H'_c(l',\tau_2)\rangle_0}{Z_c}\biggr]\;\nonumber \\
\eea
where $\hat H'_c(l,\tau)=u_{l+1}(\tau)-u_l(\tau)$ is the charge part of $H'$.
From (\ref{eq:rho_rho_2}) it is clear that when dimer-dimer correlations $\langle {\vec S_{l+1}}\cdot{\vec S_l}(\tau_1) {\vec S_{l'+1}}\cdot{\vec S_{l'}}(\tau_2)\rangle^{(0)}$ are 
present (presumably when $T \lesssim J_0$) then a $2k_F$ component appears
in $ \langle \rho(x,\tau)\rho(x',\tau')\rangle$.  After some algebra, we reach
the final form
\bea
\label{eq:rho_rho_2_final_exact}
 \langle \rho(x,\tau)\rho(x',\tau')\rangle^{(2)}
=(J_1)^2\sum_{l,l'}\int_0^{\beta} d\tau_1 \int_0^{\tau_1}d\tau_2 \nonumber \\
\times \langle {\vec S_{l+1}}\cdot{\vec S_l}(\tau_1) {\vec S_{l'+1}}\cdot{\vec S_{l'}}(\tau_2)\rangle^{(0)}\nonumber \\
\times 
\sum_{n,m}\int\frac{d\eta}{2\pi}\frac{d\xi}{2\pi} e^{i\eta(x-an)} e^{i\xi(x'-am)}\nonumber \\ 
\times e^{-\frac{1}{2}\sum_q |M_q|^2(\eta^2+\xi^2)} e^{-\eta \xi \sum_q |M_q|^2 e^{i\theta_q(n,m;\tau,\tau')}} \nonumber \\
\times \prod_{k>0}^{k_{\rm max}} |M_k|^44(1-\cos(ka))^2e^{-2(\tau_1-\tau_2)\hbar \omega_k}\nonumber \\
\times \biggl(-1 + \prod_{k'>0}^{k_{\rm max}}\Bigl(1+h(k')\Bigr)\biggr)\;,
\eea
where
\bea
h(k)=|M_k|^2 e^{-4\tau_2\hbar\omega_k}\Biggl[\eta^2 \Biggl(1-\frac{e^{2\tau_2\hbar \omega_k}R(2\tilde n)}{1-\cos(ka)}\Biggr)
\nonumber \\ +2\eta\xi \Biggl(\cos\Bigl(ka(l-l')\Bigr)-\frac{e^{2\tau_2\hbar \omega_k}R(\tilde n+\tilde m)}{1-\cos(ka)}\Biggr)\nonumber \\
+\xi^2 \Biggl(1-\frac{e^{2\tau_2\hbar \omega_k}R(2\tilde m)}{1-\cos(ka)}\Biggr)\Biggr],
\eea
with 
\be
\tilde n=n+i\omega_k \tau/(ka),
\ee
\be
\tilde m = m+i\omega_k\tau'/(ka),
\ee
and
\bea
R(s)=\cos\Bigl(ka(2l'+2-s)\Bigr)+\cos\Bigl(ka(2l'+1-s)\Bigr)\nonumber \\
+\cos\Bigl(ka(2l'-s)\Bigr).\hspace{.8 cm}
\eea

\section{Computing ${\rm Im}\chi_i^{R,0}(k,\omega)$}
\label{app:FT}

${\rm Im}\chi_i^{R,0}(k,\omega)$ is computed by making use of the Fourier expansion of $\rho^{\rm eff}_0(x,\tau)$, Eq.~\eqref{eq:rho_0_expr}, and the formula \eqref{eq:chi_tau}.  Consider first the Fourier transform to momentum space:
\be
\label{eq:rho_rho_k}
\int_{-\infty}^\infty \!\!\! dx   e^{-ik x} \langle \rho^{\rm eff}_0(x,\tau)\rho^{\rm eff}_0(0,0)\rangle
=\langle \rho^{\rm eff}_0(k,\tau)\rho^{\rm eff}_0(-k,0)\rangle,
\ee
where translational invariance was used.  The Fourier decomposition of $ \rho^{\rm eff}_0(k,\tau)$
is readily obtained by making use of Eq.~\eqref{eq:rho_0_expr}, the relation 
$u_0(x)/a=\sqrt{2}\theta_c(x)/\pi$, and the representation of $u_0(x)$ given in Eq.~\ref{eq:u_l}:
\be
\rho^{\rm eff}_0(k,\tau)=-\frac{ik}{a}\sqrt{\frac{\hbar L}{2m\omega_k}}\left(a_ke^{-\omega_k \tau} + a^\dagger_{-k} e^{\omega_k \tau}\right),
\ee
where we have implicitly converted the discrete $k$ sums to integrals and $L$ is the length of the
system. It is easily verified that this has the right units to give $\langle \rho^{\rm eff}_0(k,\tau)\rho^{\rm eff}_0(-k,0)\rangle$ in \eqref{eq:rho_rho_k} the correct dimensions of inverse length.  Only the expectation values of the cross terms $a_ka_k^\dagger$ and $a_{-k}^\dagger a_{-k}$ are nonzero,
giving
\bea  
\langle \rho^{\rm eff}_0(k,\tau)\rho^{\rm eff}_0(-k,0)\rangle = -\frac{k^2}{a^2}\frac{\hbar L}{2m\omega_k}
\hspace{2.5 cm} \nonumber \\
\times \left[e^{-\omega_k \tau}(1+n_B(k))+e^{\omega_k \tau}n_B(-k)\right],
\eea
where $n_B(k)=\langle a^\dagger_ka_k\rangle=(e^{\beta \omega_k}-1)^{-1}$ is the boson
occupation factor.  Returning to the
expression \eqref{eq:chi_tau} and evaluating the $\tau$ integral we find
\be
\chi^{0}_i(k,\omega_n)=\frac{k^2}{a^2}\frac{\hbar L}{2m\omega_k}\left[\frac{-1}{i\omega_n-\omega_k}+\frac{1}{i\omega_n+\omega_k}\right],
\ee
which upon the analytic continuation to real frequencies $i\omega_n\to \omega+i\eta$ leads directly
to Eq.~\eqref{eq:chi_R0_final}.

\section{Expressions for $I_{2k_F}(K_c,K_s,v_s/v_c)$ and $I_{4k_F}(K_c)$}
\label{app:AB}

The functions $I_{2k_F}(K_c,K_s,v_s/v_c)$ and $I_{4k_F}(K_c)$ defined in Eq.~\eqref{eq:AB_ratio} are given by
\be
\label{eq:I_2kF}
I_{2k_F}(K_c,K_s,v_s/v_c)=\int_0^\infty \frac{d\omega}{T}\frac{\Xi_{2k_F}(K_c,K_s,v_s/v_c,\omega/T)}{
\sinh^2(\omega/2T)},
\ee
and 
\be
\label{eq:I_4kF}
I_{4k_F}(K_c)=\int_0^\infty \frac{d\omega}{T}\frac{\Xi_{4k_F}(K_c,\omega/T)}{
\sinh^2(\omega/2T)},
\ee
where
\bea
\Xi_{2k_F}(K_c,K_s,v_s/v_c,\omega/T)=\Biggl({\rm Im}\int_{-\infty}^\infty \!\!\! d\tilde x  \int_{0}^\infty \!\!\! d\tilde t e^{i\frac{\omega}{\pi T}\tilde t}\nonumber \\
\times {\rm Im}\Biggl[\frac{1}{[\sinh(\tilde x- \tilde t)\sinh(\tilde x +\tilde t)]^\frac{K_c}{2}}
\hspace{3 cm}\nonumber \\
 \times \frac{1}{[\sinh\left(\tilde xv_c/v_s-\tilde t\right)\sinh(\tilde xv_c/v_s+\tilde t)]^\frac{K_s}{2}}\Biggr]\Biggr)^2,\;\;\;
 \eea
and
\bea
\Xi_{4k_F}(K_c,\omega/T)=\Biggl({\rm Im}\int_{-\infty}^\infty \!\!\! d\tilde x  \int_{0}^\infty \!\!\! d\tilde t e^{i\frac{\omega}{\pi T}\tilde t}\nonumber \\
\times {\rm Im}\Biggl[\frac{1}{[\sinh(\tilde x- \tilde t)\sinh(\tilde x +\tilde t)]^{2K_c}}
\Biggr]\Biggr)^2.\;\;\;
 \eea


\end{document}